\documentclass[aps,prl,twocolumn,superscriptaddress]{revtex4-2}
\usepackage{graphics,graphicx, array, afterpage}
\usepackage{qcircuit,amsmath}
\usepackage{grffile}
\usepackage[export]{adjustbox}
\usepackage{lineno}
\usepackage{xcolor}
\usepackage{longtable}
\usepackage{braket}
\usepackage{array}
\usepackage{multirow}
\usepackage{enumerate}
\usepackage{amsmath}
\usepackage{amssymb}
\usepackage{amsthm,times}

\usepackage[colorlinks, citecolor=blue]{hyperref}
\hypersetup{linkcolor=magenta,urlcolor=blue,citecolor=blue,pdfstartview={FitH},urlcolor=blue}

\usepackage[latin9]{inputenc}
\usepackage{color}
\usepackage{tabularx}
\usepackage{algorithm}
\usepackage{algpseudocode}
\usepackage{graphbox}
\usepackage{amsfonts}
\usepackage{dcolumn}
\usepackage{bm}
\usepackage{epstopdf}

\begin{document}

\title{Extracting Quantum Many-Body Scarred Eigenstates with Matrix Product States}

\author{Shun-Yao Zhang}
\thanks{These authors contributed equally to this work.}
\affiliation{Center for Quantum Information, IIIS, Tsinghua University, Beijing 100084, People's Republic of China}
\author{Dong Yuan}
\thanks{These authors contributed equally to this work.}
\affiliation{Center for Quantum Information, IIIS, Tsinghua University, Beijing 100084, People's Republic of China}

\author{Thomas Iadecola}
\email{iadecola@iastate.edu}
\affiliation{Department of Physics and Astronomy, Iowa State University, Ames, Iowa 50011, USA}
\affiliation{Ames National Laboratory, Ames, Iowa 50011, USA}

\author{Shenglong Xu}
\email{slxu@tamu.edu}
\affiliation{Department of Physics and Astronomy, Texas A\&M University,
College Station, Texas 77843, USA}

\author{Dong-Ling Deng}
\email{dldeng@tsinghua.edu.cn}
\affiliation{Center for Quantum Information, IIIS, Tsinghua University, Beijing 100084, People's Republic of China}
\affiliation{Shanghai Qi Zhi Institute, 41st Floor, AI Tower, No. 701 Yunjin Road, Xuhui District, Shanghai 200232, China}

\begin{abstract}

Quantum many-body scarred systems host nonthermal excited eigenstates immersed in a sea of thermal ones. In cases where exact expressions for these special eigenstates are not known, it is computationally demanding to distinguish them from their exponentially many thermal neighbors. We propose a matrix-product-state (MPS) algorithm, dubbed DMRG-S, to extract such states at system sizes far beyond the scope of exact diagonalization. Using this technique, we obtain scarred eigenstates in Rydberg-blockaded chains of up to $80$ sites and perform a finite-size scaling study to address the lingering question of the stability for the N\'eel state revivals in the thermodynamic limit.
Our method also provides a systematic way to obtain \textit{exact} MPS representations for scarred eigenstates near the target energy without \textit{a priori} knowledge. In particular, we find several new scarred eigenstates with exact MPS representations in kinetically constrained spin and clock models. The combination of numerical and analytical investigations in our work provides a new methodology for future studies of quantum many-body scars.
\end{abstract}

\maketitle
\begin{figure}
\hspace*{-0.45\textwidth}
\includegraphics[width=0.95\linewidth]{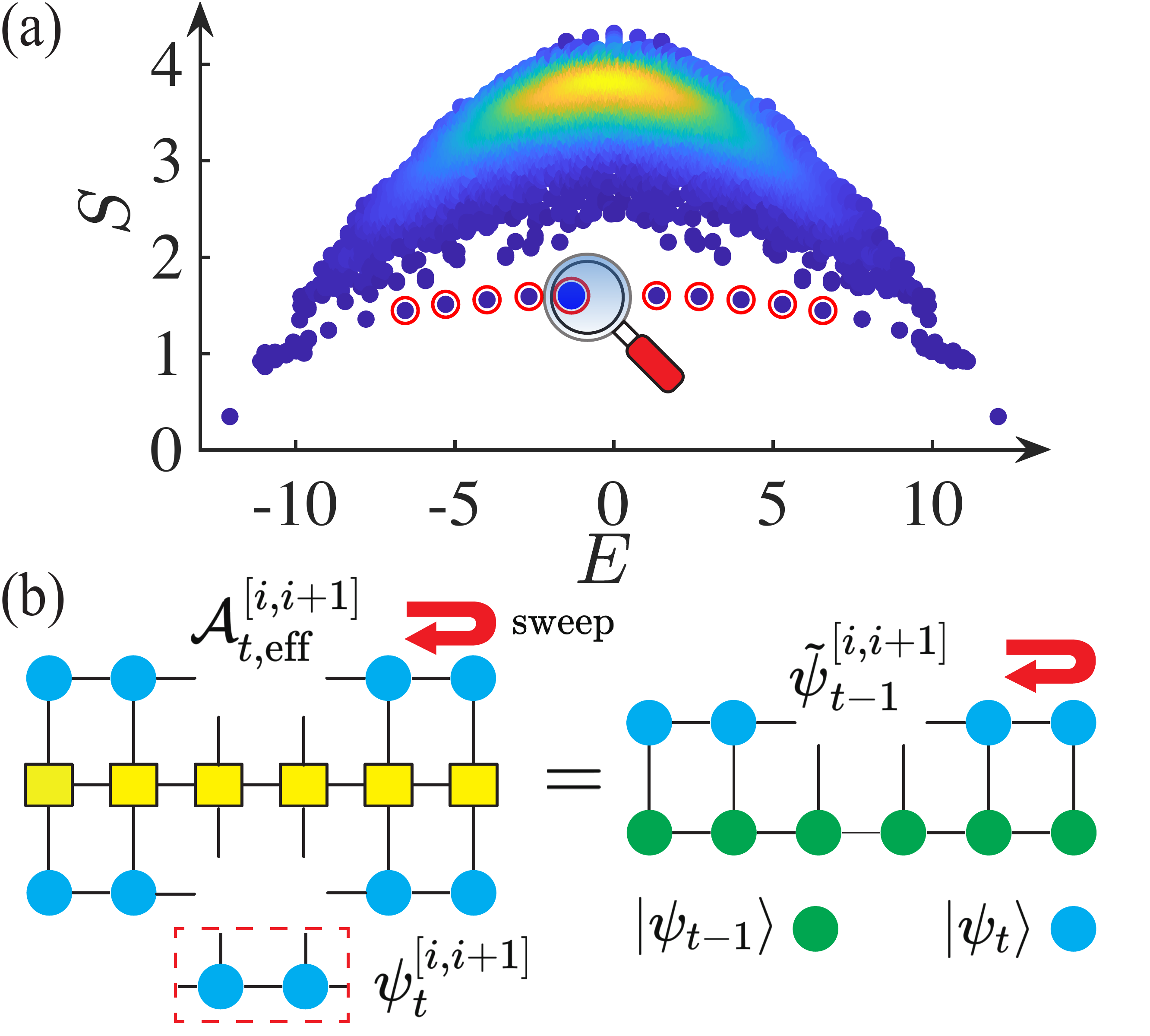} 
\caption{Schematic illustration of the DMRG-S algorithm for extracting quantum many-body scars  with matrix product states. (a) Density plot showing the bipartite entanglement entropy $S$ versus energy eigenvalue $E$ for the PXP model. DMRG-S effectively serves as a magnifier to discover low-entanglement scar states within a target energy window. (b) Schematic of the variational procedure for obtaining the updated matrix product state $|\psi_t\rangle$ (blue circles) from $|\psi_{t-1}\rangle$ (green circles) by locally solving the linear equation $\mathcal{A}_{t,\text{eff}}^{[i,i+1]} \psi_t^{[i,i+1]} = \tilde{\psi}_{t-1}^{[i,i+1]}$, where $\mathcal{A}_t=(H-\xi_t)^2$ (yellow blocks). }
\label{Algorithm Illustration}
\end{figure}

 
Quantum many-body scars (QMBS) appear in many-body systems with weak ergodicity breaking~\cite{Bernien2017Probing,Serbyn2021quantum,Moudgalya2021Quantum,Chandran2022ScarReview}. These anomalous scarred eigenstates violate the eigenstate thermalization hypothesis \cite{Deutsch1991Quantum,Srednicki1994Chaos,Deutsch2018Eigenstate,Rigol2008Thermalization,Kim2014testing}, yet only comprise a vanishing fraction of the Hilbert space, as opposed to the strong ergodicity breaking in integrable \cite{Sutherland2004beautiful} or many-body localized systems \cite{Nandkishore2015Manybody,Abanin2019Colloquium}. Typical many-body scarred eigenstates possess sub-volume-law entanglement entropy, and are immersed in a sea of thermal eigenstates [see Fig.~\ref{Algorithm Illustration}(a)]. Many models exist in which a set of scarred eigenstates can be calculated analytically~\cite{Shiraishi2017Systematic,Moudgalya2018Exact,Moudgalya2018Entanglement,Choi2019emergent,lin2019exact,Schecter2019Weak,ok2019topological,Chattopadhyay2020quantum,Moudgalya2020eta,Lee2020Exact,Langlett2022rainbow,Langlett2021Hilbert,Schindler2022Exact}, but there are other examples in which their appearance remains mysterious. For instance, experiments in Rydberg-atom quantum simulators realizing the ``PXP model"~\cite{Bernien2017Probing,Bluvstein2021Controlling} found evidence of QMBS in the dynamics of an initial N\'eel state, which exhibited coherent revivals for unexpectedly long time owing to its high overlap with a tower of scarred eigenstates. Motivated by these experiments, a flurry of theoretical and experimental works have emerged to explain the rich properties of these special eigenstates \cite{Turner2018weak,Turner2018quantum,Khemani2019Signatures,Ho2019periodic,Choi2019emergent,Michailidis2020Slow,Surace2020Lattice,Magnifico2020Real,Iadecola2019Quantum,Turner2021Correspondence,yuan2022quantum,Pan2022Composite,Omiya2023Quantum} and find other models hosting many-body scars \cite{Ho2019periodic,bull2019systematic,hudomal2020quantum,Scherg2021Observing,Desaules2021Proposal,Su2022Observation,Banerjee2021Quantum,Biswas2022Scars,Desaules2022Weak,Desaules2022Prominent,Zhang2022Many}.

In such cases without exact analytical expressions for the scarred eigenstates, their existence can be confirmed by full diagonalization of the Hamiltonian followed by a calculation of some diagnostics, e.g.~the entanglement entropy, across the whole spectrum. The exponential computational cost of exact diagonalization (ED) poses a substantial challenge to faithfully addressing the fate of QMBS in the thermodynamic limit. Examples of questions that are difficult to address using ED include the ultimate fate of periodic revivals for the N\'eel state in the PXP model \cite{Bernien2017Probing,Choi2019emergent} and the robustness of scarred eigenstates under various perturbations \cite{Lin2020Slow,Surace2021Exact,Shem2021Fate,Huang2021Stability}. 
Matters can be further complicated by the fact that highly excited eigenstates of many-body Hamiltonians can have exponentially large degeneracy in the presence of certain symmetries \cite{Schecter2018Many,karle2021area,buijsman2022number,Banerjee2021Quantum,Biswas2022Scars}. This renders the task of finding scars using ED methods extremely difficult in general.

Scarred eigenstates in one dimension often have entanglement entropy scaling at most logarithmically with the system size~\cite{Moudgalya2018Entanglement,Turner2018quantum,Schecter2019Weak,Iadecola2020Quantum,Chattopadhyay2020quantum}, suggesting that they could be described using matrix product state (MPS) representations at system sizes inaccessible to ED~\cite{Orus2014Practical,Cirac2021Matrix}. 
In this paper, we propose an MPS-based algorithm to extract quantum many-body scarred eigenstates with high accuracy (see Fig.~ \ref{Algorithm Illustration} for a pictorial illustration). To demonstrate its power, we compute the tower of scarred eigenstates for system sizes up to $L=80$ in the PXP model \cite{Turner2018weak,Turner2018quantum} and a deformation thereof~\cite{Khemani2019Signatures,Choi2019emergent}. With a detailed finite-size scaling study, we find that the coherent revivals of the N\'eel state vanish in the thermodynamic limit in the PXP model, whereas they remain stable in the deformed PXP model.

Moreover, previous analytical studies have shown that highly excited scarred eigenstates in several models possess exact MPS representations \cite{Moudgalya2018Entanglement,lin2019exact,Chattopadhyay2020quantum,Moudgalya2020Large,karle2021area}, while the constructions of these scars are model-specific and lack generalizability. In contrast, our method provides a systematic way to find exact MPS representations for QMBS in generic Hamiltonians, without \textit{a priori} knowledge. We use our algorithm to discover several new zero-energy scarred eigenstates with exact MPS representations in the kinetically constrained clock \cite{bull2019systematic} and higher-spin PXP models \cite{Ho2019periodic}. 
We also find \textit{a posteriori} analytical derivations for these scars that apply to a wide variety of kinetically constrained models.

\textit{DMRG-S Algorithm.}-- Our algorithm is inspired by the density matrix renormalization group (DMRG) method \cite{white1992density,schollwock2011density}, which has been widely used to obtain modestly entangled ground states of low-dimensional Hamiltonians. In the past few years, DMRG methods relying on the MPS formalism have been generalized to obtain highly excited eigenstates of many-body localized systems \cite{Khemani2016Obtaining,Kennes2016Entanglement,Lim2016Many,yu2017finding,Serbyn2016Power}. In this work, we modify and improve the  shift-invert technique \cite{Luitz2015Manybody,yu2017finding,Serbyn2016Power} to be amenable for calculating scarred eigenstates. Below, we dub the algorithm DMRG-S, where ``S" stands for ``scars".

The algorithm is based on the intuition that repeatedly applying the inverse operator $(H-\xi)^{-2}$ (more robust and efficient in convergence compared to $(H-\xi)^{-1}$~\cite{SuppMaterials}) to an initial state $\ket{\psi_0}$ eventually yields an eigenstate of $H$ with energy $\xi$, provided $\ket{\psi_0}$ has overlap with this eigenstate. In practice, we define $\ket{\psi_0}$ to be an MPS and consider the sequence of states $\ket{\psi_t}=\mathcal N^{-1}\mathcal A_t^{-1}\ket{\psi_{t-1}}$, where $\mathcal A_t=(H-\xi_t)^{2}$ and $\mathcal N$ is a normalization factor (We describe an update procedure for $\xi_t$ below). The state $\ket{\psi_t}$ is taken to be an MPS with bond dimension $\chi\leq \chi_{\rm max}$. Restricting $\chi_{\rm max}$ to relatively small values effectively serves as a filter for states with low entanglement entropy. In the iteration step $t$, we circumvent the difficulty of calculating the inverse operator $\mathcal A_t^{-1}$ by variationally optimizing $\ket{\psi_t}$ such that $\braket{\psi_t|\mathcal A_{t}|\psi_t}=\mathcal N^{-1}\braket{\psi_t|\psi_{t-1}}$. This approach has the advantage that $\mathcal A_t$ can be expressed as a matrix product operator. The optimization can be implemented by locally solving the linear equation
\begin{equation}
     \mathcal{A}_{t,\text{eff}}^{[i,i+1]} \psi_t^{[i,i+1]}=   \tilde{\psi}_{t-1}^{[i,i+1]},
\label{eq:linear}
\end{equation}
where $ \mathcal{A}_{t,\text{eff}}^{[i,i+1]} $ is the local ``effective Hamiltonian" for $\mathcal{A}_t$,  $\psi_t^{[i,i+1]}$ is the local tensor of $|\psi_{t} \rangle$ to be updated, and $\tilde{\psi}_{t-1}^{[i,i+1]}$ is the environment tensor of the overlap  $\langle \psi_t |\psi_{t-1}\rangle$ [see Fig.~\ref{Algorithm Illustration}(b)].
The optimized $\psi_t^{[i,i+1]}$ is substituted back into $\ket{\psi_t}$, which is then brought to the canonical form via singular value decomposition.
We perform the local optimization on each pair of sites $[i,i+1]$ sweeping back and forth, similar to the two-site DMRG sweep procedure \cite{white1992density, schollwock2011density}. During the iterations, we monitor the energy variance $\sigma_H^2 = \langle H^2 \rangle - \langle H \rangle^2$ of $\ket{\psi_t}$, which vanishes if and only if $\ket{\psi_t}$ is an eigenstate. 
Initially we set $\xi_0$ within the target energy window $[E-\Delta E, E+\Delta E]$, which may not contain the energy of the initial state $\ket{\psi_0}$. After a few iterations, if $\sigma_H^2$ reaches a relatively small value (less than $10^{-3}$), we then begin to update $\xi_t=\langle \psi_{t}|H|\psi_{t} \rangle$ during each iteration. The update of the energy shift $\xi_t$ is crucial for the convergence if we do not \textit{a priori} know the precise locations of scars in the energy spectrum~\cite{SuppMaterials}. These two stages correspond to the slow and fast decay regions shown in Fig.~\ref{Exact MPS}(b).
Eventually we expect $|\psi_{t} \rangle$  to converge, i.e.~$\lim_{t\to \infty} |\langle \psi_{t-1}|\psi_{t} \rangle|^2 = 1$, and approach to an eigenstate with energy close to the target one.

\begin{figure*}
\hspace*{-0.98\textwidth}
\includegraphics[width=0.98\linewidth]{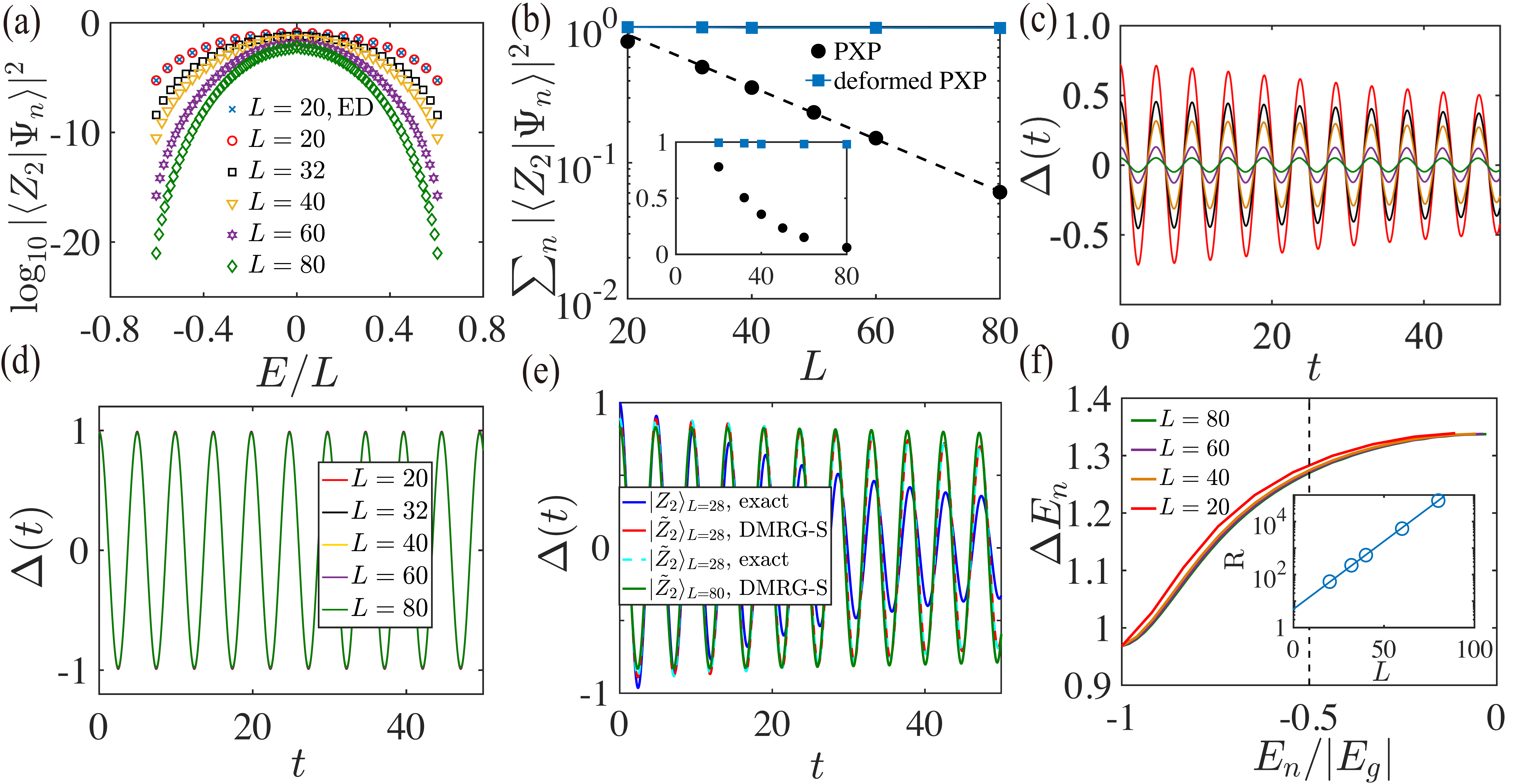}
\caption{Numerical results for the tower of scars in the (deformed) PXP model. (a) Overlap between the N\'eel state $|Z_2\rangle$ and each scarred eigenstate of the PXP model for different $L$, all obtained by DMRG-S except points marked by crosses. (b) Finite-size scaling for the total overlap between $|Z_2\rangle$ and the $L+1$ scarred eigenstates of the (deformed) PXP model. The inset displays data on the linear scale. (c) Dynamics of the staggered magnetization density $\Delta$ within the scarred subspace constructed by DMRG-S ($\mathbb P = \sum^L_{n=0}\ket{\Psi_n}\bra{\Psi_n}$), for the PXP model. (d) The same dynamics for the deformed PXP model. (e) Observable dynamics ${\Delta}(t)$ of $\ket{\tilde{Z}_2}$ for the PXP model, which exhibits more stable revivals than $\ket{Z_2}$ (blue). ${\Delta}(t)$ dynamics of $\ket{\tilde{Z}_2}$ computed by using the DMRG-S eigenenergies (red) and by exact Hamiltonian evolution (cyan dashed) agree well with each other.
(f) Energy spacings $\Delta E_n$ between adjacent scars as a function of the normalized eigenenergy $E_n/|E_{g}|$ for the PXP model. The inset shows that the ratio $R$ increases exponentially with $L$.}
\label{Numerical Results}
\end{figure*}

\textit{Tower of scars in PXP models.}--
The PXP Hamiltonian is the effective Hamiltonian for a chain of spins satisfying the Rydberg blockade constraint, which forbids configurations containing $\ket{\uparrow}_i\ket{\uparrow}_{i+1}$ due to strong nearest-neighbor interactions \cite{Jaksch2000Fast,Lukin2001Dipole,Bernien2017Probing}. It is given by  $H_{PXP}=\sum_i P_i X_{i+1} P_{i+2}$, where $P_i = (1-Z_i)/2 $ projects onto $\ket{\downarrow}_i$ and $X_i,Z_i$ are Pauli matrices on site $i$.
$H_{PXP}$ is nonintegrable according to studies of its level statistics, and yet hosts a tower of scars supporting the periodic revival dynamics of the N\'eel state $\ket{Z_2}=\ket{\uparrow\downarrow\uparrow\cdots\downarrow}$ \cite{Turner2018weak,Turner2018quantum}. Numerical simulations of these dynamics observe that the revivals have a decaying envelope, begging the question of whether they persist at late time in the thermodynamic limit. Ref.~\cite{Choi2019emergent} found that adding a term
$ \delta H_2= -  h_2 \sum_i  P_{i-1} X_i P_{i+i}(Z_{i-2}+Z_{i+2}) $ with $h_2 = 1/2 - 1/ \sqrt{5} \approx 0.053$ makes the periodic revivals nearly perfect due to the emergence of an approximate $su(2)$ algebra. Here, we benchmark the DMRG-S algorithm by computing the tower of scarred eigenstates in the PXP model and its deformation by $\delta H_2$.

We initialize the algorithm in the state $\ket{\psi_0}=\ket{Z_2}$, which has predominant overlap with the $L+1$-dimensional tower of scarred eigenstates $\{\ket{\Psi_n}\}_{n=0}^{L}$ within corresponding energy windows.  
During the iterations, we set $\chi_{\rm max}=1200$ to reach the desired accuracy due to the logarithmic scaling of subsystem entanglement entropy \cite{Turner2018weak,Turner2018quantum} and the periodic boundary conditions. As shown in Fig.~\ref{Numerical Results}, DMRG-S successfully extracts the tower of scars in the PXP model up to $L=80$. The average energy variance $\sigma_H^2$ is less than $10^{-6}$ \cite{SuppMaterials}. To verify that these MPSs indeed capture the scar tower of the PXP Hamiltonian,  we calculate their overlap with $|Z_2\rangle$ [Fig.~\ref{Numerical Results}(a)], and their bipartite entanglement entropy \cite{SuppMaterials} for different $L$.
Our results yield smooth curves as a function of energy and agree with ED for small system sizes except for a few scars that accidentally hybridize with thermal eigenstates \cite{Turner2018quantum,Iadecola2019Quantum}, which are further addressed in~\cite{hybrid_footnote,SuppMaterials}. 

We now investigate the quench dynamics of $\ket{Z_2}$ by finite-size scaling beyond the scope of ED using DMRG-S states up to $L=80$. First, we compute the total overlap between $\ket{Z_2}$ and $\{\ket{\Psi_n}\}_{n=0}^{L}$ [Fig.~\ref{Numerical Results}(b)], and find that $\sum^L_{n=0} |\langle Z_2 | \Psi_n\rangle|^2$ decays exponentially with $L$ for the PXP model. In contrast, this quantity remains near unity for the deformed PXP model. The dashed line in Fig.~\ref{Numerical Results}(b) ($y = e^{-0.044 L + 0.739}$) is obtained from linear regression with $R^2 \approx 0.9996$. 
To further probe the revivals, we evaluate the dynamics of the staggered magnetization density $\Delta=[\sum^L_{i=1}(-1)^{i+1}Z_i]/L$ within the scarred subspace constructed by DMRG-S: $\Delta (t) = \langle Z_2 | \mathbb{P}  e^{i H t}  \Delta e^{-i H t} \mathbb{P} | Z_2 \rangle \approx \sum_{n,m=0}^L e^{i(E_n - E_m)t} \langle Z_2 | \Psi_n \rangle  \langle \Psi_n | \Delta | \Psi_m \rangle \langle \Psi_m | Z_2 \rangle$, where $\mathbb P = \sum^L_{n=0}\ket{\Psi_n}\bra{\Psi_n}$, $\{E_n\}_{n=0}^L$ and $\{\ket{\Psi_n}\}_{n=0}^L$ are scarred eigenenergies and eigenstates obtained via DMRG-S \cite{Delta_t_footnote}.
$\Delta (t)$ characterizes the late-time non-thermal observable dynamics after the local relaxation time (the infinite-temperature value of $\Delta$ is zero).
Fig.~\ref{Numerical Results}(c) and (d) display $\Delta(t)$ as a function of time for different $L$ in the PXP and deformed PXP models, respectively.
We find that the oscillation amplitude shrinks with increasing $L$ for the PXP model but remains unaltered for the deformed case, consistent with our results for the total $|Z_2\rangle$ overlap.

Furthermore, we evaluate the observable dynamics of the deformed $Z_2$ state $\ket{\tilde{Z}_2} = \mathbb{P} \ket{Z_2} / \sqrt{\langle Z_2 | \mathbb{P} | Z_2 \rangle}$ constructed by DMRG-S (which has logarithmic entanglement \cite{SuppMaterials}) in the PXP model. As shown in Fig.~\ref{Numerical Results}(e), oscillations of $\Delta(t) = \langle \tilde{Z}_2 | e^{i H t} \Delta e^{-i H t} | \tilde{Z}_2 \rangle$ become more stable as system size increases, suggesting the robustness of the periodic revivals for $\ket{\tilde{Z}_2}$ in the thermodynamic limit.
To illustrate this phenomenon, we calculate the energy spacings $\Delta E_n$ between adjacent scars as a function of $E_n/|E_g|$ [Fig.~\ref{Numerical Results}(f)], where $E_{g}$ is the ground state energy and $n=0,1,\cdots,L/2$ label the scars from the spectrum boundary to center. Notably, we find that $\Delta E_n$ approaches an $L$- and $n$-independent constant  near the center of spectrum ($E=0$).
Furthermore, inspired by Fig.~\ref{Numerical Results}(a), we compute the ratio $R = \sum_{n \in C} |\langle Z_2| \Psi_n \rangle|^2 / \sum_{n \in B} |\langle Z_2| \Psi_n \rangle|^2$, where the vertical dashed line $E_n/|E_g|=-0.5$ in Fig.~\ref{Numerical Results}(f) separates $\ket{\Psi_n}$ belonging to the spectrum center ($C$) or boundary ($B$). As shown in the inset of Fig.~\ref{Numerical Results}(f), $R$ increases exponentially with the system size. Combining these two observations, we deduce that the equidistant scars near the center of spectrum dominate the revival dynamics of $\ket{\tilde{Z}_2}$ as $L$ increases, resulting in the more stable oscillations observed in Fig.~\ref{Numerical Results}(e).

To sum up, for the PXP model the coherent revivals of the N\'eel state vanish in the thermodynamic limit due to its exponentially small overlap with the scarred subspace, whereas the revivals remain stable in the deformed case. Nevertheless, our results demonstrate that one can stabilize the revivals in the original PXP model by initializing in a modestly entangled state like $\ket{\tilde{Z}_2}$. The DMRG-S algorithm provides a convenient method to construct such states \cite{SuppMaterials}.

\textit{Exact MPS representations for QMBS.}-- 
Apart from the ability to extract QMBS at system sizes beyond the scope of ED, our algorithm also opens up a promising avenue to naturally obtain the exact MPS representations for certain QMBS. Several exact scars have been discovered in previous analytical studies, such as the $E=\pm \sqrt{2}$ scars $\ket{\Gamma_{12}},\ket{\Gamma_{21}}$ and the $E=0$ scars $\ket{\Gamma_{11}},\ket{\Gamma_{22}}$ in the spin-$1/2$ PXP model \cite{lin2019exact}, and the $E=0$ scar in the deformed one-dimensional cluster model~\cite{ok2019topological}.
We first benchmark our algorithm by recovering the above known examples. We run the DMRG-S algorithm for about 200 random initial states and select the converged MPS with smallest variance $\sigma_H^2$. 
During the optimization we fix $\chi_{\rm max} = 10$. As shown in Fig.~\ref{Exact MPS}(a), even though the fidelity $f=|\langle \psi_t | \Psi_\text{exact} \rangle |$ is initially exceedingly small ($\sim 10^{-6}$), DMRG-S can extract these exact scarred eigenstates to high precision within 100 iterations. We stress that our algorithm is not hindered by the exponentially large degeneracy in the $E=0$ eigensubspace~\cite{Schecter2018Many,karle2021area,buijsman2022number} and does not utilize any \textit{a priori} knowledge. Thus it can be applied to generic many-body Hamiltonians in any target energy window.

\begin{figure}
\hspace*{-0.5\textwidth}
\includegraphics[width=1.0\linewidth]{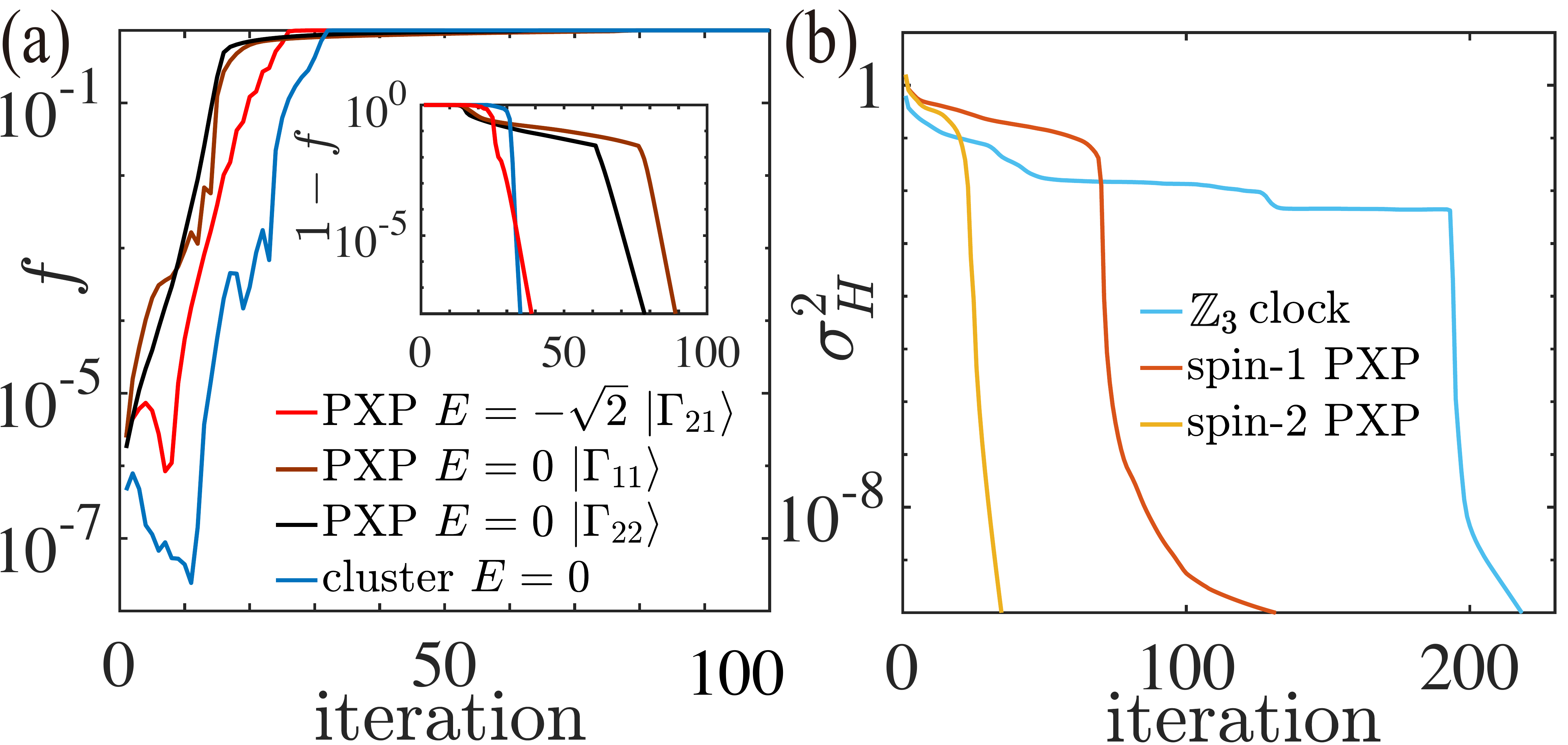}
\caption{ (a) Fidelity $f=|\langle \psi_t | \Psi_\text{exact} \rangle |$ between the optimized MPSs and exact scars from the spin-$1/2$ PXP model \cite{lin2019exact} and the deformed one-dimensional cluster model \cite{ok2019topological} as a function of iteration number. The inset shows the infidelity $1-f$. (b) Energy variance $\sigma^2_H$ of the optimized MPSs for the $\mathbb{Z}_3$ clock and spin-$1$ and $2$ PXP models as a function of iteration number. 
}
\label{Exact MPS}
\end{figure}

Indeed, in the kinetically constrained clock model \cite{bull2019systematic} and higher-spin PXP models \cite{Ho2019periodic}, we discover several $E=0$ scarred eigenstates with exact MPS representations that have not been reported in previous literature. As shown in Fig.~\ref{Exact MPS}(b), the energy variance of the optimized MPSs in the corresponding models converges to very small values ($\sim 10 ^{-10}$) within 200 iterations. We further apply the singular value decomposition to compress their bond dimensions, typically to $\chi=2$ for the open boundary cases, then continue the optimization until convergence again. Careful analysis of the bulk tensors on each site yields the expressions reported below. We write the MPS representations as $|\Psi\rangle = \sum_{\sigma} \mathrm{Tr}\left(A^{[\sigma_1]}_1 A^{[\sigma_2]}_2 \cdots A^{[\sigma_L]}_L \right) \ket{\sigma_1\sigma_2\cdots\sigma_L} $ for periodic boundary conditions, where $\sigma=\sigma_1\sigma_2\cdots\sigma_L$ denotes the physical index of each site. We define the following $2\times 2$ matrices: 
\begin{equation}
B = 
\left(\begin{array}{cc}
1 & 0 \\
0 & 0
\end{array}\right), \quad
D =  
\left(
\begin{array}{cc}
1 & 1 \\
-1 & -1 
\end{array}\right).
\label{Eq:B_D_mat}
\end{equation}
These matrices are related to those found in the numerical calculations by appropriate MPS gauge transformations~\cite{Orus2014Practical,Cirac2021Matrix}.

For the kinetically constrained $\mathbb{Z}_N$ clock model~\cite{bull2019systematic} $H_{\text{clock}}= \sum_i P_{i-1} C_{i} P_{i+1}$, the local Hilbert space is spanned by $N$ states $\{\ket{0},\ket{1},\cdots,\ket{N-1}\}$. Here, $P_i=\ket{0}_i\bra{0}_i$ forbids creating excitations (i.e., basis states besides $\ket{0}$) on neighboring sites, and $\mathcal{U}_i=\exp(-i C_i)=\sum_{n=0}^{N-1} \ket{n+1}_i\bra{n}_i$ cyclically permutes basis states on site $i$ (we define $\ket{N}\equiv \ket{0}$). A translationally invariant highly excited eigenstate $\ket{\Psi}_c$ with $E=0$ can be constructed using $A^{[0]} = B$, $A^{[1],[2],\cdots,[N-1]} = D$. In~\cite{SuppMaterials} we show that $P_{i-1} C_{i} P_{i+1} \ket{\Psi}_c = 0$, $\forall\ i$. We further observe that this MPS is nothing but the equal-weight superposition of all computational basis states allowed by the constraints $\ket{\Psi}_c=\sum_{\text{allowed}\ \sigma} \ket{\sigma_1\sigma_2\cdots\sigma_L}$. 

The spin-$s$ PXP models~\cite{Ho2019periodic} are defined by $H_{PXP}=\sum_i P_{i-1} S^x_{i} P_{i+1}$, where the local Hilbert space is spanned by $2s+1$ states $\{\ket{-s},\ket{-s+1},\cdots,\ket{s-1},\ket{s}\}$. $P_i=\ket{-s}_i\bra{-s}_i$, and $S^x_{i}$ is the spin-$s$ generator of rotations around the $x$-axis.
When $s$ is an integer, a translationally invariant scarred eigenstate $\ket{\Psi}_s$ with $E=0$ can be expressed as
\begin{equation}
A^{[-s]} = B, \quad A^{[-s+2k-1]} = 0, \quad A^{[-s+2k]} = a_k D,   
\label{Eq:PXP_s_MPS}   
\end{equation}
where $k=1,2,3,\cdots,s$, and $a_k = \langle m_z = -s+2k | m_x = 0\rangle/\langle m_z = -s |  m_x = 0\rangle$~\cite{a_k_footnote}. 
Similarly, $P_{i-1} S^x_{i} P_{i+1} \ket{\Psi}_s = 0$, $\forall\ i$~\cite{SuppMaterials}.
$\ket{\Psi}_s$ also takes a simple form in the computational basis,
\begin{equation}
\ket{\Psi}_s=\sum_{\text{allowed}\ \sigma} \left[\prod_{k=1}^s (a_k)^{\# \text{ of }-s+2k \text{ in }\sigma}\right] \ket{\sigma_1\sigma_2\cdots\sigma_L},
\end{equation}
where the allowed computational basis states contain only local states $\{\ket{-s+2k}\}_{k=0}^s$ and the additional prefactors count the number of $\ket{-s+2k}$ states appearing in $\ket{\sigma_1\sigma_2\cdots\sigma_L}$.

The above exact scars can be analytically derived as follows. Consider Hamiltonians of the form $H= \sum_i P_{i-1} h_i P_{i+1}$, where the local Hilbert space is spanned by the bases $\{\ket{0},\ket{1},\cdots,\ket{d-1}\}$ and $P_i=\ket{0}_i\bra{0}_i$. We define the projector onto the global constrained Hilbert space as $P = \prod_i (I - \tilde{P}_i \tilde{P}_{i+1})$, where $\tilde{P_i} = I - P_i$. 
If the single-site operator $h_i$ has a zero mode $\ket{\phi_i}$ (e.g.~$\sum_{n=0}^{N-1}\ket{n}_i$ for the clock model, and $\ket{m_x=0}_i$ for the PXP models of integer spins), the product state $\ket{\Phi}=\prod_i \ket{\phi_i}$ is a zero-energy eigenstate of $H$. While this state does not satisfy the global constraint defined by $P$, the projected state $P\ket{\Phi}$ does, and in fact remains a zero-energy eigenstate since $[P,H]=0$. Since $P$ can be expressed as a matrix product operator with bond dimension $\chi = 2$, the zero-energy scarred eigenstate $P \ket{\Phi}$ becomes an MPS with bond dimension $\chi = 2$. Explicit calculations \cite{SuppMaterials} yield the $2\times 2$ matrices in Eq.~\eqref{Eq:B_D_mat} and the coefficients in Eq.~\eqref{Eq:PXP_s_MPS}.  We stress that this construction is different from the embedding construction of Ref.~\cite{Shiraishi2017Systematic}, where the embedded scarred eigenstates are annihilated by certain local projectors $P_i$ rather than the local operators $h_i$.

{\it Conclusion.}-- In summary, we have introduced the DMRG-S algorithm to accurately extract quantum many-body scarred eigenstates. This method can access system sizes far beyond the scope of ED and assist analytical studies in discovering exact MPS representations of new scars for generic Hamiltonians. It also sheds light on other open questions about QMBS, such as their robustness under various types of perturbations~\cite{Lin2020Slow,Surace2021Exact,Shem2021Fate,Huang2021Stability}. The analytical construction of exact scars inspired by our numerical results provides a different mechanism for scar states in models with local kinetic constraints. 
The synergy between numerical calculations and analytical investigations in our work establishes a promising framework for future studies on quantum many-body scars.

The DMRG-S algorithm is implemented based on the ITensor library~\cite{fishman2020itensor} in Julia programming language. 
The source code for the numerical calculations is accessible online~\cite{Source_code}.

\begin{acknowledgments}
We acknowledge helpful discussions with Luming Duan and He-Ran Wang. We especially thank Zlatko Papi\'c for comments on an earlier version of the manuscript. S.-Y.~Z., D.~Y., and D.-L.~D. acknowledge support from the National Natural Science Foundation of China (Grants No. 12075128),  Tsinghua University, and Shanghai Qi Zhi Institute. T.I. acknowledges support from the National Science Foundation through Grant No.~DMR-2143635.
\end{acknowledgments}

\bibliographystyle{apsrev4-1-title}
\bibliography{DengQAIGroup,QMBS}

\begin{thebibliography}{78}
\bibitem [{Del()}]{Delta_t_footnote}%
  \BibitemOpen
  \href@noop {} {}\bibinfo {note} {As long as the eigenenergy variances are
  small enough for the DMRG-S states, the $\Delta(t)$ dynamics computed by the
  DMRG-S eigenenergies will agree with the exact Hamiltonian evolution. We
  explicitly benchmark the observable dynamics for small system sizes in Fig.
  2(e) and \cite{SuppMaterials}.}\BibitemShut {Stop}%
\end{thebibliography}

\clearpage
\onecolumngrid
\makeatletter

\setcounter{MaxMatrixCols}{10}

\setcounter{figure}{0}
\renewcommand{\thefigure}{S\@arabic\c@figure}
\setcounter{equation}{0} \makeatletter
\renewcommand \theequation{S\@arabic\c@equation}
\renewcommand \thetable{S\@arabic\c@table}

\begin{center} 
	{\large \bf Supplementary Materials for: Extracting Quantum Many-Body Scarred Eigenstates with Matrix Product States}
\end{center}

\section{Analytical Proofs for Exact MPS Representations of QMBS}
In this section, we provide the analytical proofs for the exact matrix product state (MPS) representations of quantum many-body scars (QMBS) found by our DMRG-S algorithm.

General one dimensional (1D) quantum states can be written as the MPS form
\begin{equation}
|\Psi \rangle  = \sum_{\sigma} \mathrm{Tr}\left( A_1^{[\sigma_1]}  A_2^{[\sigma_2]} \cdots  A_L^{[\sigma_L]} \right) | \sigma_1\sigma_2 \cdots \sigma_L \rangle,
\label{PBC_MPS}
\end{equation}
in the periodic boundary condition (PBC), and 
\begin{equation}
|\Psi \rangle  = \sum_{\sigma} v_l A_1^{[\sigma_1]}  A_2^{[\sigma_2]} \cdots  A_L^{[\sigma_L]} v_r^T | \sigma_1\sigma_2 \cdots \sigma_L \rangle,
\label{OBC_MPS}
\end{equation}
in the open boundary condition (OBC). $\sigma = \sigma_1 \sigma_2 \cdots \sigma_L$ denotes the physical index of each site, in other words, $\{|\sigma_i\rangle\}$ are the local bases on site $i$. For example, $\sigma_i=\uparrow, \downarrow$ for spin-$1/2$ systems. $\{A_i^{[\sigma_i]}\}$ are $\chi_{i-1} \times \chi_{i}$ matrices for all $\sigma_i$. $\{\chi_i \}$ are called bond dimensions, which upper bound the entanglement entropy $S\sim O[\log(\chi_i)]$ of $\ket{\Psi}$ at the cut $i,i+1$ \cite{Orus2014Practical,Cirac2021Matrix}. $v_l,v_r$ are left and right boundary vectors. 
The product of $A_i^{[\sigma_i]}$ matrices provides the complex probability amplitude of the computational basis $| \sigma_1 \sigma_2 \cdots \sigma_L \rangle$ in $\ket{\Psi}$.  Similarly, general 1D Hamiltonians can be written as the matrix product operator (MPO) form
\begin{equation}
H = \sum_{\sigma, \sigma^\prime} H_1^{[\sigma_1], [\sigma_1^\prime]}  H_2^{[\sigma_2],[\sigma_2^\prime]} \cdots  H_L^{[\sigma_L],[\sigma_L^\prime]} | \sigma_1 \sigma_2 \cdots \sigma_L \rangle \langle \sigma_1^\prime \sigma_2^\prime  \cdots \sigma_L ^\prime|
\end{equation}
where $\{H_i^{[\sigma_i], [\sigma_i^\prime]}\}$ are $\gamma_{i-1} \times \gamma_{i}$ matrices, and $\sigma_i$,  $\sigma_i^\prime$ are physical indices on the site $i$.

\subsection{Known Examples of Exact QMBS}
We first briefly review the known examples of exact QMBS shown in Fig. 3(a) of the main text. In the spin-$1/2$ PXP model of OBC, Ref. \cite{lin2019exact} has shown that the following MPSs 
\begin{equation}
\left|\Gamma_{\alpha\beta}\right\rangle=\sum_{\sigma} v_{\alpha} B^{[\sigma_{1}]} C^{[\sigma_{2}]} \cdots B^{[\sigma_{L-1}]} C^{[\sigma_{L}]} v_{\beta}^T\left|\sigma_{1} \ldots \sigma_{L}\right\rangle,
\end{equation}
where $\sigma_i = 0(\downarrow),1(\uparrow)$, $\alpha,\beta=1,2$, and 
\begin{equation}
\begin{array}{ll}
B^{[0]}=\left(\begin{array}{ccc}
1 & 0 & 0 \\
0 & 1 & 0
\end{array}\right),  
B^{[1]}=\sqrt{2}\left(\begin{array}{ccc}
0 & 0 & 0 \\
1 & 0 & 1
\end{array}\right), 
C^{[0]}=\left(\begin{array}{cc}
0 & -1 \\
1 & 0 \\
0 & 0
\end{array}\right),  
C^{[1]}=\sqrt{2}\left(\begin{array}{cc}
1 & 0 \\
0 & 0 \\
-1 & 0
\end{array}\right),
v_1 = (1,1),  v_2 =(1,-1),

\end{array}
\end{equation}
are exact scarred eigenstates with $E=0$ for $|\Gamma_{11}\rangle$ and  $|\Gamma_{22}\rangle$, $E=\sqrt{2}$ for $|\Gamma_{12}\rangle$, and $E=-\sqrt{2}$ for $|\Gamma_{21} \rangle$.

We have also considered the deformed one-dimensional cluster model~\cite{ok2019topological} with OBC,   
\begin{align}
&H(\beta) = \sum_{i=1}^{L} \alpha_i Q_i(\beta),
\quad \alpha_i = \alpha+(-1)^{i}, \nonumber\\ 
&Q_1(\beta) =  e^{-\beta Z_{1} Z_{2}} - X_1,
\quad Q_L(\beta) = e^{-\beta Z_{L-1} Z_{L}} - X_L, \nonumber\\
&Q_i(\beta) =  e^{-\beta(Z_{i-1}Z_i+Z_{i} Z_{i+1})} - X_i, \quad \forall i \neq 1,L,
\end{align}
where $0<|\alpha|<1$, $\beta \neq 0$ (specifically, we take $\alpha=0.3$, $\beta=0.5$).
There exists an $E=0$ excited scarred eigenstate for $H(\beta)$
\begin{equation}
|\text{scar}(\beta)\rangle=G(\beta) \bigotimes_{i=1}^L|+\rangle_{i},
\end{equation}
where $G(\beta)=\exp \left(\frac{\beta}{2} \sum_{i=1}^{L-1} Z_{i} Z_{i+1}\right)$, $\ket{+}=(\ket{0}+\ket{1})/\sqrt{2}$.
Note that $|\text{scar}(\beta)\rangle$ is also the ground state of the Hamiltonian
$\widehat{H}(\beta) =\sum_{i=1}^{L} |\alpha_i| Q_i(\beta)$,
which results in the area-law entanglement of $|\text{scar}(\beta)\rangle$.

\subsection{General Schemes for Zero-energy Scars in Kinetically Constrained Hamiltonians}
Consider general kinetically constrained Hamiltonians of the following form
\begin{equation}
H = \sum_i P_i h_{i} P_{i+1},
\end{equation}
where the local Hilbert space is spanned by $\{\ket{0},\ket{1},\cdots,\ket{d-1}\}$ with the dimension $d$, $P_i=\ket{0}_i\bra{0}_i$. We define the projector towards the global constrained Hilbert space as 
\begin{equation}
P = \prod_i (I - \tilde{P}_i \tilde{P}_{i+1}), \quad \tilde{P_i} = I - P_i.
\end{equation}
One can verify that $[P,H]=0$.

If there exists a zero mode $\ket{\phi_i}$ for the single-site operator $h_i$, we define the product state
\begin{equation}
\ket{\Phi}=\prod_i \ket{\phi_i},
\end{equation}
which is annihilated by each term of the Hamiltonian, thus $H \ket{\Phi} = 0$.
Since $H$ commutes with the projector $P$, the projection of the product state $\ket{\Phi}$ into the constrained Hilbert space will become a zero-energy scarred eigenstate $H P \ket{\Phi} = 0$.

Now $P \ket{\Phi}$ is no longer a product state because of the projector $P$. To obtain the explicit MPS representation of $P \ket{\Phi}$, we construct the MPO representation for the projector $P$ as follows. We break each term of $P$ into a product of two vectors
\begin{equation}
(I - \tilde{P}_i \tilde{P}_{i+1}) = 
\begin{pmatrix}
I & \tilde{P}_i
\end{pmatrix}
\begin{pmatrix}
I \\ -\tilde{P}_{i+1}
\end{pmatrix}.
\end{equation}
Then we have
\begin{equation}
P = \mathrm{Tr} \left(\prod_i 
\begin{pmatrix}
I &  \tilde{P}_i \\
-\tilde{P}_i & -\tilde{P}_i
\end{pmatrix}
\right),
\end{equation}
which is in an explicit MPO form with bond dimension $\chi = 2$. 
Then $P \ket{\Phi}$, obtained by contracting $P$ with a product state $\ket{\Phi}$, can be expressed in an MPS with bond dimension $\chi = 2$:
\begin{equation}
P\ket{\Phi} =  \mathrm{Tr} \left(\prod_i 
\begin{pmatrix}
\ket{\phi_i} &  \tilde{P}_i \ket{\phi_i} \\
-\tilde{P}_i \ket{\phi_i} & -\tilde{P}_i \ket{\phi_i}
\end{pmatrix}
\right) = \sum_{\sigma}\mathrm{Tr} \left(\prod_i A^{[\sigma_i]}_i \right) | \sigma_1\sigma_2 \cdots \sigma_L \rangle,
\end{equation}
where 
\begin{equation}
A^{[0]}_i =\langle 0|\phi_i\rangle\begin{pmatrix}
1 & 0 \\
0 & 0
\end{pmatrix}, \quad A^{[\sigma_i > 0]}_i =  \langle \sigma_i |\phi_i \rangle \begin{pmatrix}
1 &  1\\
-1 & -1
\end{pmatrix}.    
\end{equation}

As mentioned in the main text, we define the following $2\times 2$ matrices
\begin{equation}
B = 
\left(\begin{array}{cc}
1 & 0 \\
0 & 0
\end{array}\right), \quad
D =  
\left(
\begin{array}{cc}
1 & 1 \\
-1 & -1 
\end{array}\right),
\end{equation}
satisfying $D^2=0$, and $B D B=BBB$. Below for the $\mathbb{Z}_N$ clock model with kinetic constraints and integer spin-$s$ PXP models, we specifically calculate the coefficients before $B,D$ matrices and analyze the properties of the scarred eigenstates.

\subsection{Kinetically Constrained Clock Models}
For the $\mathbb{Z}_N$ clock model with kinetic constraints \cite{bull2019systematic}, there exist $N$ bases on each site $i$, $\{\ket{0},\ket{1},\cdots,\ket{N-1}\}$. The ``\textit{PCP}" Hamiltonian reads
\begin{equation}
    H_{\text{clock}} = \sum_{i} P_{i-1} C_i P_{i+1}, \quad P_i=\ket{0}_i\bra{0}_i, \quad \mathcal{U}_i=\exp(-i C_i)=\sum_{n=0}^{N-1} \ket{n+1}_i\bra{n}_i.
\label{clock_Ham}
\end{equation}
Each site $i$ precesses around the clock bases if both its neighbors are in $\ket{0}$, otherwise its rotation remains frozen. Note that $\mathcal{U}_i^{N}=\mathbb{I}$, thus eigenvalues of $\mathcal{U}_i$ have the forms of $\omega^{k_n}$, where $\omega = \exp(i 2 \pi /N)$ and $k_n$ are arbitrary integers. Since $C_i \propto i \ln \mathcal{U}_i$, the expressions of $C_i$ are not unique. Here we set the values of $\{k_n\}_{n=0}^{N-1}$ as 
\begin{equation}
k_n = \begin{cases}
-\frac{N-1}{2}, -\frac{N-1}{2}+1, \cdots, 0, \cdots, \frac{N-1}{2}-1, \frac{N-1}{2} \quad \text{for odd } N \\
-(\frac{N}{2}-1), -(\frac{N}{2}-1)+1, \cdots, 0, \cdots, \frac{N}{2}-1, \frac{N}{2} \quad \text{for even } N.
\end{cases} 
\label{k_n_values}
\end{equation}
Notice that values of $\{k_n\}_{n=0}^{N-1}$ for even $N$ taken here are slightly different from those in \cite{bull2019systematic} since we preserve the clock period $T$ $(\mathcal{U}_i^{T}=\mathbb{I})$ to be $T=N$ for both even and odd $N$. 
Now $C_i$ can be expressed as
\begin{equation}
    C_i = -\frac{2\pi}{N}\sum_{n=0}^{N-1} k_n \ket{\psi_n}_i \bra{\psi_n}_i, \quad \ket{\psi_n}_i = \frac{1}{\sqrt{N}}\sum_{j=0}^{N-1} \omega^{-k_n j } \ket{j}_i,
\end{equation}
with $\{\ket{\psi_n}_i\}_{n=0}^{N-1}$ being the eigenstates of $\mathcal{U}_i$ and $C_i$. Below we list the $N = 2, 3, 4$ $C_i$ matrices as examples.

\begin{equation}
C_{i} = \frac{\pi}{2}
    \left(\begin{array}{cc}
     -1 & 1 \\
     1  & -1
\end{array}\right)_i,
\quad
C_{i} = i\frac{2\pi}{3\sqrt{3}}
    \left(\begin{array}{ccc}
     0 & -1 & 1 \\
     1 & 0 & -1 \\
     -1 & 1 & 0
\end{array}\right)_i,
\quad
C_{i} = \frac{\pi}{4}
    \left(\begin{array}{cccc}
     -1 & 1-i & -1 & 1+i\\
     1+i & -1 & 1-i & -1\\
     -1 & 1+i & -1 & 1-i \\
     1-i & -1 & 1+i & -1
\end{array}\right)_i.
\label{C_i_234}
\end{equation}

As mentioned in the main text, a translationally invariant QMBS $\ket{\Psi}_c$ with $E=0$ can be expressed as 
\begin{equation}
    A^{[0]} = B, \quad A^{[1],[2],\cdots,[N-1]} = D
\end{equation}
in PBC Eq. \eqref{PBC_MPS}, and with 
\begin{equation}
    v_{l} = v_{r} = \left(1, 0\right)
\label{boundary_vec}
\end{equation}
in OBC Eq.~\eqref{OBC_MPS}. First, the kinetic constraints are satisfied since $A^{[\sigma_i]}A^{[\sigma_{i+1}]}=0$, for $\sigma_i,\sigma_{i+1} \neq 0$. Second, since $A^{[0]} A^{[\sigma_i]} A^{[0]} = A^{[0]} A^{[0]} A^{[0]}$, $\forall \sigma_i = 0,1,2,\cdots,N-1$, the coefficients of the computational bases $\ket{\sigma_1\cdots, 0,\sigma_i, 0, \cdots \sigma_{L}}$ in $\ket{\Psi}_c$ are same for all $\sigma_i$. 
Observed from Eq. \eqref{C_i_234} (also rigorously proved later), the summation for each row of the $C_i$ matrix is zero. Combining these two results together, we deduce that 
$A^{[0]} \times \left(\sum_{\sigma_i'=0}^{N-1} (C_i)^{\sigma_i \sigma_i'} A^{[\sigma_i']} \right) \times A^{[0]} = 0$, $\forall \sigma_i$,
thus $P_{i-1} C_{i} P_{i+1} \ket{\Psi}_c = 0$, $\forall i$.
For the boundary terms, we have $v_l A_1^{[\sigma_1]} A_2^{[0]} = v_l A_1^{[0]} A_2^{[0]}$ and $A_{L-1}^{[0]} A_L^{[\sigma_L]} v_r^T = A_{L-1}^{[0]} A_L^{[0]} v_r^T$, $\forall \sigma_1,  \sigma_L= 0,1,2,\cdots,N-1$, thus $ C_{1} P_{2} \ket{\Psi}_c = 0 $ and $P_{L-1} C_{L} \ket{\Psi}_c = 0$ in the OBC case.

According to the analysis above, we further deduce that superposition coefficients of \textit{all} the computational bases in $\ket{\Psi}_c$ are the same, so this MPS is nothing but the equal-weight superposition state of all the allowed computational bases in the constrained Hilbert space 
\begin{equation}
    \ket{\Psi}_c=\sum_{\text{allowed}\ \sigma} \ket{\sigma_1\sigma_2\cdots\sigma_L}
\label{Psi_c_com}
\end{equation}
up to a normalization factor, where the allowed computational basis $\ket{\sigma}$ means that each two neighboring sites $\ket{\sigma_i},\ket{\sigma_{i+1}}$ should contain at least one $\ket{0}$ basis. It is quite surprising that such a simply-constructed state in computational bases can be the $E=0$ highly excited scarred eigenstates of $H_{\text{clock}}$ and expressed as an MPS with finite bond dimension $\chi=2$, which demonstrates the power of our DMRG-S algorithm. Notice that there exist other $B,D$ matrices satisfying the conditions $D^2=0$ and $BDB=BBB$, but they all represent the same physical state above in computational bases.

Next, we rigorously show that the summation for each row or column of the $C_i$ matrix is zero (in the following we omit the subscript $i$ for convenience). To begin with, $\ket{\psi_n} = \frac{1}{\sqrt{N}} \sum_{j=0}^{N-1} \omega^{-k_n j } \ket{j}$ is the eigenstate of $\mathcal{U}$ with eigenvalue $\omega^{k_n}$ [see the values of $k_n$ in Eq. \eqref{k_n_values}], such that the $C$ matrix can be decomposed into
\begin{equation}
    C = -\frac{2\pi}{N} E \left(\begin{array}{cccc}
     k_0 &  &  & \\
      & k_1 &  & \\
      &  & \ddots &  \\
      &  &  & k_{N-1}
\end{array}\right) E^{-1}, \quad E_{j n}= \frac{1}{\sqrt{N}} \omega^{-j k_n}, \quad (E^{-1})_{j n}= \frac{1}{\sqrt{N}} \omega^{n k_j},
\end{equation}
where $j,n = 0, 1, \cdots, N-1$. Now we can explicitly calculate the the summation for the $i$-th row of $C$ as
\begin{equation}
    \sum_{i=0}^{N-1} C_{ij} = -\frac{2 \pi}{N} \sum_{i,l=0}^{N-1} E_{i l} \times k_l \times (E^{-1})_{l j} = -\frac{2 \pi}{N^2} \sum_{i,l=0}^{N-1} k_l \omega^{-(i-j) k_l} = 0.
\end{equation}
The summation for the $j$-th column is also zero through similar calculations.

Besides, we analyze the symmetry properties of the exact QMBS $\ket{\Psi}_c$. First, the ``\textit{PCP}" Hamiltonian possesses the inversion symmetry $I: i \to L-i+1$. The operator $I$ transforms the MPS representations in PBC and OBC by
\begin{equation}
I|\Psi \rangle_c  = \sum_{\sigma} \mathrm{Tr}\left( A^{[\sigma_1]}_I  A^{[\sigma_2]}_I \cdots  A^{[\sigma_L]}_I \right) | \sigma_1\sigma_2 \cdots \sigma_L \rangle,
\end{equation}
\begin{equation}
I|\Psi \rangle_c  = \sum_{\sigma} v_r A^{[\sigma_1]}_I  A^{[\sigma_2]}_I \cdots  A^{[\sigma_L]}_I v_l^T | \sigma_1\sigma_2 \cdots \sigma_L \rangle,
\end{equation}
where $A^{[\sigma_i]}_I = \left(A^{[\sigma_i]}\right)^T $ and we have omitted the site indices due to the translational invariance of $\ket{\Psi}_c$. The $2 \times 2$ $z$-Pauli matrix $\sigma^z$ will give us a gauge transformation restoring the original MPS representations, which proves that $I\ket{\Psi}_c = \ket{\Psi}_c$.
\begin{equation}
\sigma^z \left(A^{[\sigma_i]}\right)^T \sigma^z = A^{[\sigma_i]}, \quad v_r \sigma^z = v_l, \quad \sigma^z v_l^T = v_r^T
\label{gauge_trans}
\end{equation}

Second, given the $\{k_n\}_{n=0}^{N-1}$ values in Eq. \eqref{k_n_values}, for odd $N$, $H_{\text{clock}}$ has the particle-hole (charge conjugation) symmetry $\{H_{\text{clock}}, \mathcal{C}_{p h}\}=0$ \cite{bull2019systematic}, leading to the $E \leftrightarrow -E$ symmetry in the spectrum, where
\begin{equation}
    \mathcal{C}_{p h} = \prod_{i=1}^{L} \mathcal{C}_i, \quad \mathcal{C}_i = 
    \left(\begin{array}{ccccc}
      1 & 0 & \cdots & 0 & 0\\
      0 & 0 & \cdots & 0 & 1\\
      0 & 0 & \cdots & 1 & 0\\
      0 & \cdots & 1 & 0 & 0\\
      \vdots & \vdots &\vdots &\vdots & \vdots \\
      0 & 1 & 0 & \cdots & 0 
\end{array}\right)_i.
\end{equation}
Note that for even $N$, although the spectrum of $H_{\text{clock}}$ is not symmetric with respect to zero, $E=0$ still corresponds to highly excited energy. The operator $\mathcal{C}_i$ keeps $\ket{0}_i$ invariant and exchanges $\ket{\sigma_i}_i$ with $\ket{N-\sigma_i}_i$ for $\sigma_i = 1, 2, \cdots, N-1$. Since $A^{[1],[2],\cdots,[N-1]} = C$, we deduce that $\mathcal{C}_{p h} \ket{\Psi}_c = \ket{\Psi}_c$ for odd $N$.

\subsection{PXP Models for Integer Spins}
As for general PXP models of spin-$s$ (numerically shown to be nonintegrable and quantum chaotic by level statistics study \cite{Ho2019periodic}), each site $i$ contains $2s+1$ bases $\{\ket{-s},\ket{-s+1},\cdots,\ket{s-1},\ket{s}\}$. The Hamiltonian reads
\begin{equation}
    H_{PXP}=\sum_i P_{i-1} S^x_{i} P_{i+1}, \quad P_i=\ket{-s}_i\bra{-s}_i, \quad \bra{s,m_z \pm 1}S^x\ket{s,m_z}=\sqrt{(s\pm m_z +1)(s \mp m_z)}/2,
\end{equation}
where $S^x_{i}$ is the $x$-angular momentum operator of spin-$s$. When $s$ is an integer, we start by presenting the concrete examples of $s=1,2$. The $S^x_{i}$ operators for $s=1,2$ are
\begin{equation}
S^x_{i} = \frac{1}{\sqrt{2}}
    \left(\begin{array}{ccc}
     0 & 1 & 0\\
     1 & 0 & 1\\
     0 & 1 & 0
\end{array}\right)_i,
\quad
S^x_{i} = 
    \left(\begin{array}{ccccc}
     0 & 1 & 0 & 0 & 0 \\
     1 & 0 & \sqrt{\frac{3}{2}} & 0 & 0 \\
     0 & \sqrt{\frac{3}{2}} & 0 & \sqrt{\frac{3}{2}} & 0 \\
     0 & 0 & \sqrt{\frac{3}{2}} & 0 & 1 \\
     0 & 0 & 0 & 1 & 0 
\end{array}\right)_i.
\end{equation}
Actually since the eigenvalues of $S^x_i$ for integer spins and $C_i$ matrices for $\mathbb{Z}_N$ clock models with odd $N$ are the same, by performing the basis transformation, the \textit{PCP} Hamiltonian Eq. \eqref{clock_Ham} can be expressed in the spin bases, where $C_i$ becomes $S^x_i$ and consequently the kinetic constraint $P_i$ in Eq. \eqref{clock_Ham} will be changed \cite{bull2019systematic}. The following similar expressions of exact QMBS also reflect the close relations between these two models.

A translationally invariant $E=0$ scarred eigenstate $\ket{\Psi}_1$ for $s=1$ can be expressed in the MPS form as
\begin{equation}
A^{[-1]}=B,\quad A^{[0]} = 0,\quad A^{[1]}=-D
\end{equation}
in PBC, and with the same boundary vectors Eq. \eqref{boundary_vec} in OBC.

The proof for its eigenstate property follows the similar way as the clock model. The kinetic constraints are satisfied since $A^{[\sigma_i]}A^{[\sigma_{i+1}]}=0$, for $\sigma_i,\sigma_{i+1} \neq -1$. Because $A^{[0]}=0$ and $A^{[-1]} A^{[-1]} A^{[-1]} + A^{[-1]} A^{[1]} A^{[-1]}=0$, by noticing the matrix elements in each row of spin-$1$ $S^x_i$, we deduce that 
$A^{[-1]} \times \left(\sum_{\sigma_i'=-1}^{1} (S^x_i)^{\sigma_i \sigma_i'} A^{[\sigma_i']}\right) \times A^{[-1]} = 0$, $\forall \sigma_i$,
thus $P_{i-1} S^x_{i} P_{i+1} \ket{\Psi}_1 = 0$, $\forall i$. The boundary terms in OBC annihilates $\ket{\Psi}_1$ through the same approach.
Then according to the matrix elements in spin-$2$ $S^x_i$, we require that $A^{[\sigma_i]}A^{[\sigma_{i+1}]}=0$, for $\sigma_i,\sigma_{i+1} \neq -2$, $A^{[-1]} = 0$, $A^{[-2]} A^{[-2]} A^{[-2]} + \sqrt{\frac{3}{2}} A^{[-2]} A^{[0]} A^{[-2]}=0$, $\sqrt{\frac{3}{2}} A^{[-2]} A^{[0]} A^{[-2]} + A^{[-2]} A^{[2]} A^{[-2]} = 0$ and $A^{[1]} = 0$, such that an $E=0$ exact QMBS $\ket{\Psi}_2$ for the $s=2$ PXP model should be like 
\begin{equation}
A^{[-2]}=B,\quad A^{[-1]}=0,\quad A^{[0]}=-\sqrt{\frac{2}{3}} D,\quad A^{[1]} = 0,\quad A^{[2]} = D    
\end{equation}
in PBC, and with the boundary vectors Eq. \eqref{boundary_vec} in OBC.

Through this approach, we can generalize the MPS representations above to arbitrary integer spins. According to the matrix elements in spin-$s$ $S^x_i$ $\bra{s,m_z\pm 1}S^x\ket{s,m_z}=\sqrt{(s\pm m_z +1)(s \mp m_z)}/2$, we obtain the following recurrence relations
\begin{equation}
    \sqrt{(2 k-1)(s-k+1)} A^{[-s]} A^{[-s+2 k-2]} A^{[-s]}+ \sqrt{k (2s-2k+1)} A^{[-s]} A^{[-s+2 k]} A^{[-s]} = 0 \quad k=1,2,\cdots,s,
\label{recur_1}
\end{equation}
\begin{equation}
    A^{[-s+1]}=A^{[-s+3]}=\cdots=A^{[s-3]}=A^{[s-1]}=0.
\label{recur_2}
\end{equation}
Therefore, we define a corresponding number sequence $\{a_k\}_{k=0}^s$ with the recurrence relation
\begin{equation}
    a_0 = 1, \quad a_k = -\sqrt{\frac{(2 k-1)(s-k+1)}{k (2s-2k+1)}} a_{k-1} \  (k=1,2\cdots,s).
\end{equation}

Together with the kinetic constraints, we deduce that an $E=0$ scarred eigenstate $\ket{\Psi}_s$ of integer spin-$s$ PXP models can be expressed as
\begin{align}
&A^{[-s]} = B, \quad A^{[-s+2k-1]}=0, \quad A^{[-s+2k]} = a_k D,  \nonumber \\
a_k &= (-1)^k \sqrt{\left[\frac{s!}{k!(s-k)!}\right] \bigg/ \left[\frac{(2s-1)!!}{(2k-1)!!(2s-2k-1)!!}\right]}, 
\end{align}
where $k=1,2,3,\cdots,s$, and $(2k-1)!!=(2k-1)\times (2k-3)\times \cdots \times 3\times 1$ is the double factorial.
The OBC cases have additional boundary vectors Eq. \eqref{boundary_vec}.

Because the nonzero bulk matrices $A^{[\sigma_i]}$ are proportional to $B$ and $D$, and the boundary vectors remain the same as the clock model, these MPSs are directly proven to possess inversion symmetry $I\ket{\Psi}_s = \ket{\Psi}_s$ through the same gauge transformation $\sigma^z$ in Eq.~\eqref{gauge_trans}.
The particle-hole symmetry operator for the PXP models now becomes $\mathcal{C}_{p h}=\prod_{i}\mathcal{C}_i$, $\mathcal{C}_i=\text{diag}(1, -1, 1, -1,\cdots,-1,1)_i$, with $\{H_{PXP}, \mathcal{C}_{p h}\}=0$. Obviously $\mathcal{C}_{p h} \ket{\Psi}_s = \ket{\Psi}_s$.

Similar to the kinetically constrained clock model, $\ket{\Psi}_s$  can be written as simple forms in computational bases 
\begin{equation}
\ket{\Psi}_s=\sum_{\text{allowed}\ \sigma} \left[\prod_{k=1}^s (a_k)^{\# \text{ of }-s+2k \text{ in }\sigma}\right] \ket{\sigma_1\sigma_2\cdots\sigma_L}.
\end{equation}
where the allowed computational basis $\ket{\sigma}$ only consists of bases $\{\ket{-s+2k}\}_{k=0}^s$ and the additional prefactors count the number ($\#$) of $\ket{-s+2k}$ bases appearing in $\ket{\sigma}$. In particular,
\begin{equation}
    \ket{\Psi}_1=\sum_{\text{allowed}\ \sigma} (-1)^{\# \text{ of 1 in } \sigma} \ket{\sigma_1\sigma_2\cdots\sigma_L}, \quad \ket{\Psi}_2=\sum_{\text{allowed}\ \sigma} (-\sqrt{\frac{2}{3}})^{\# \text{ of 0 in } \sigma} \ket{\sigma_1\sigma_2\cdots\sigma_L}.
\end{equation}

The above recursive implementations for the $E=0$ QMBS of integer spins do not apply for the half-integer cases, because the $S^x$ operator of half-integer spins does not host a local zero mode.
Nevertheless, through numerical calculations by our DMRG-S algorithm, we still find several different $E=0$ scarred eigenstates in the spin-$3/2$ PXP model with MPS representations. Their bond dimensions can be even reduced to $\chi=2$ in OBC cases. Unfortunately, unlike the situations in the clock model and PXP models for integer spins, the bulk tensors found by DMRG-S do not exhibit apparent features. The matrices $A^{[\sigma_i]}_i$ vary from site to site and the matrix elements do not share common relations. We leave the investigations about the $E=0$ scars in the PXP models of half-integer spins for future studies.

\section{Several details about the DMRG-S algorithm}
In this section we present several numerical details about the DMRG-S algorithm. In Table. \ref{alg:DMRGS}, we display the pseudo-code for the DMRG-S algorithm, which we implement based on the ITensor library \cite{fishman2020itensor} in Julia programming language. The local effective Hamiltonian $\mathcal{A}_{t,\text{eff}}^{[i,i+1]}$ and the environment tensor $\tilde{\psi}_{t-1}^{[i,i+1]}$ are constructed by contracting all the other tensors except for those on the sites $[i,i+1]$ of $\ket{\psi_t}$, as illustrated in Fig. 1(b) of the main text. 

\begin{table*}[htb]
\begin{minipage}{\linewidth}
\begin{algorithm}[H]
\caption{DMRG-S}
\begin{algorithmic}[1]
\State Set $\xi_0$ within the target energy window $[E-\Delta E, E+\Delta E]$
\State Construct the initial MPO  $\mathcal{A}_0 = (H-\xi_0)^2$
\State Construct the initial MPS $|\psi_0\rangle$
\For{each iteration $t$}
\State $|\psi_{t}\rangle = |\psi_{t-1}\rangle$ 
\For{each sweep}
\For{each site $i$ during the left or right sweep}
\State Move the orthogonality center of the MPS $|\psi_{t}\rangle$ to site $i$ by the singular value decomposition
\State Construct the effective Hamiltonian $\mathcal{A}_{t,\text{eff}}^{[i,i+1]}$ from $|\psi_{t}\rangle$, $|\psi_{t-1}\rangle$ and $\mathcal{A}_t$
\State Construct the environment tensor $\tilde{\psi}_{t-1}^{[i,i+1]}$ from $|\psi_{t}\rangle$ and $|\psi_{t-1}\rangle$
\State Solve the linear equation $\mathcal{A}_{t,\text{eff}}^{[i,i+1]} \psi_t^{[i,i+1]} = \tilde{\psi}_{t-1}^{[i,i+1]}$ with the conjugate gradient method
\State Substitute the optimized tensor $\psi_t^{[i,i+1]}$ into the MPS $|\psi_{t}\rangle$ 
\EndFor
\EndFor
\State Normalize the MPS $|\psi_t\rangle$
\State Calculate $\xi_t = \langle \psi_t | H | \psi_t \rangle$ and $\sigma_H^2(|\psi_t\rangle)  =  \langle \psi_t | H^2 | \psi_t \rangle - \xi_t ^2$
\If{$\sigma_H^2(|\psi_t\rangle) < \min\{\sigma_H^2(|\psi_{t-1}\rangle), 10^{-3}\}$}
\State Construct the MPO $\mathcal{A}_{t+1} = (H-\xi_t)^2$
\Else
\State $\mathcal{A}_{t+1} = \mathcal{A}_{t} $
\EndIf 
\EndFor
\end{algorithmic}
\end{algorithm}
\end{minipage}
\caption{Pseudo-code of the DMRG-S algorithm for calculating quantum many-body scarred eigenstates.}
\label{alg:DMRGS}
\end{table*}

\begin{figure}
\hspace*{-1.0\textwidth}
\includegraphics[width=1.0\linewidth]{ 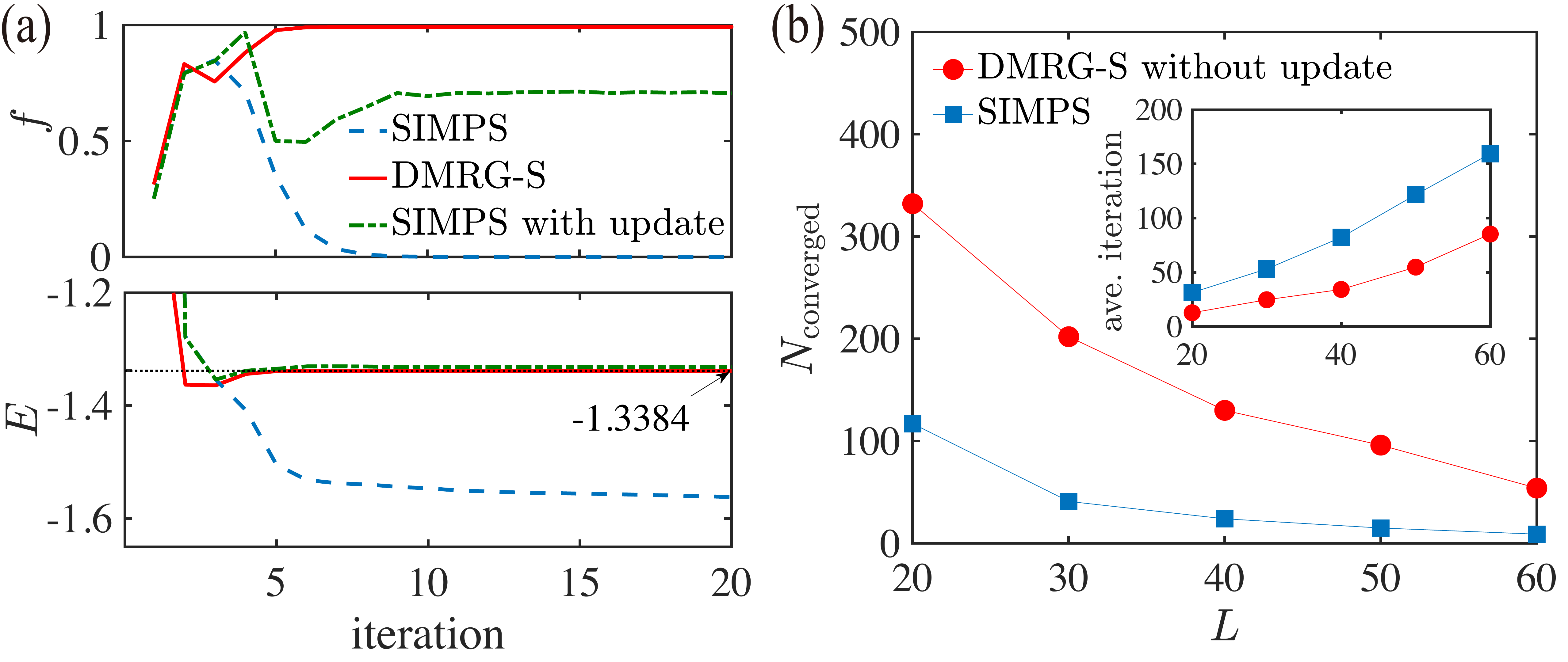} 
\caption{ Comparison for the performance of the DMRG-S and SIMPS algorithm for two different tasks. (a) Upper panel: The state fidelity $f$ between $\ket{\psi_t}$ and the $E\approx -1.33$ scar tower state in $L=30$ PXP model as a function of the iteration step. Lower panel: The energy of $\ket{\psi_t}$ as a function of the iteration step.
For the three cases, the initial energy shift $\xi_0$ is taken as $\xi_0 = -1.6$. $|\psi_0\rangle = |Z_2\rangle$ and $\chi_{\max} =200$. 
The horizontal dotted black line indicates the exact energy of the $E\approx -1.33$ scar tower state. 
(b) The number of converged ($\sigma_H^2 <10^{-6}$) MPSs by DMRG-S (without updating the energy shift $\xi_t$) and SIMPS as a function of the system size, for the task of extracting the exact scars $\ket{\Gamma_{11}}$ and $\ket{\Gamma_{22}}$ with $E=0$ in the PXP model.  
Both algorithms are initialized with 1000 random product states $|\psi_0 \rangle$ and $\xi_0 \in [-0.01, 0.01]$. $\chi_{\max}=10$. The inset shows the average iteration number for the ten fastest-converging MPSs.}
\label{fig:DMRGS_SIMPS}
\end{figure}

\textit{Comparison with existing shift-invert algorithms for MBL systems.}-- We compare our DMRG-S algorithm with existing shift-invert algorithms for many-body localized (MBL) systems, in particular, the SIMPS algorithm proposed in Ref.~\cite{yu2017finding} to find highly-excited eigenstates of MBL Hamiltonians. 
In the following, based on the properties of MBL and many-body scarred systems, we clarify the differences between the DMRG-S and SIMPS algorithm,
and numerically demonstrate the suitability and advantages of DMRG-S for extracting quantum many-body scars in Fig.~\ref{fig:DMRGS_SIMPS}.

\begin{figure}
\hspace*{-0.85\textwidth}
\includegraphics[width=0.85\linewidth]{ 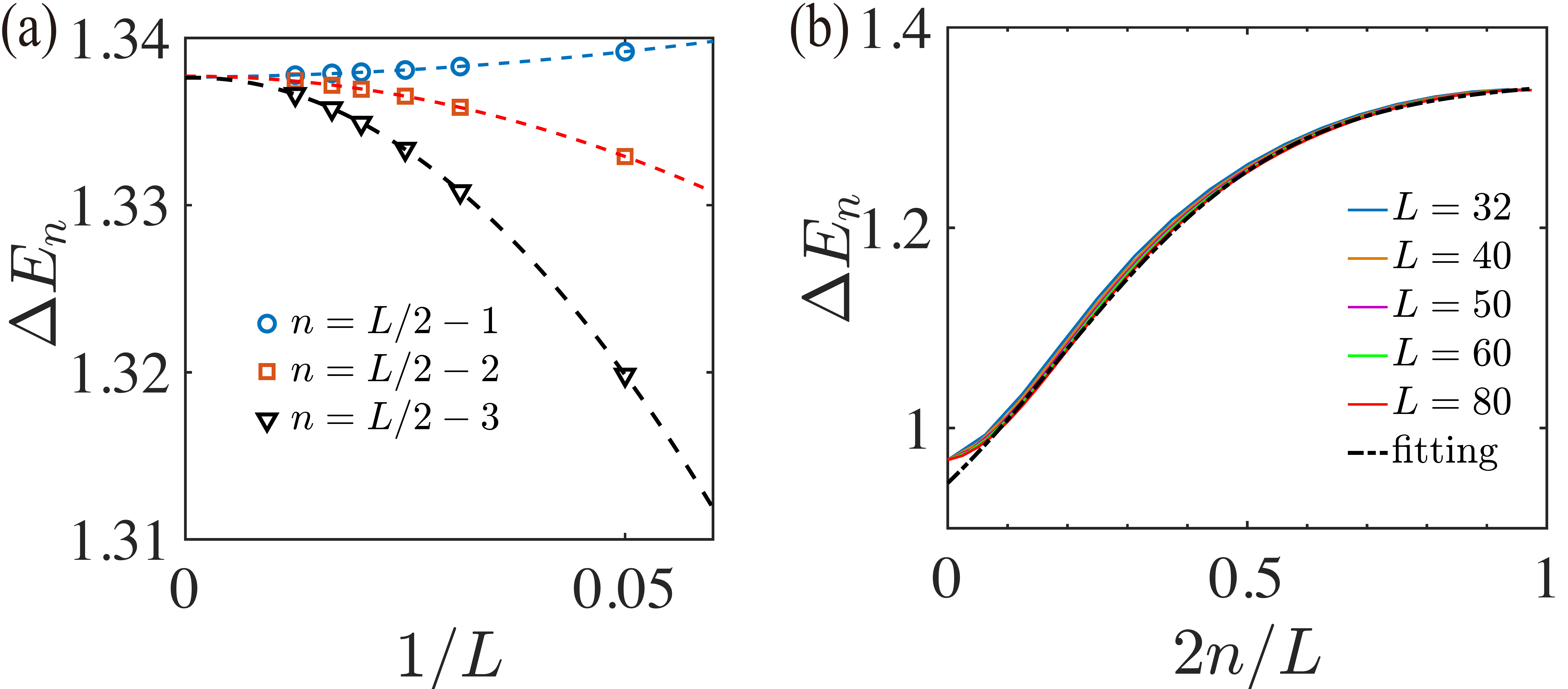} 
\caption{(a) Finite-size scaling of the energy gap $\Delta E_n = |E_{n+1}-E_n|$ between adjacent scars for $n=L/2-1,L/2-2,L/2-3$ ($n=0,1,\cdots,L/2-1$ are the indices for scars from the spectrum boundary to center) in the PXP model, which can be extrapolated to the same value $1.3377$ when $L\to \infty$. The dashed lines are obtained by the quadratic fitting. (b) $\Delta E_n$ as a function of $2n/L$ for different system sizes. The dashed line is the fitting curve with the ansatz function $y = b_0/(1 + e^{b_1-b_2 x} )+ b_3$. The fitting parameters are $(b_0,b_1,b_2,b_3)=(0.56,0.97,5.19,0.79)$. 
}
\label{fig:Delta E_n}
\end{figure}

First, in DMRG-S the energy shift $\xi$ is updated ($\xi_t=\langle \psi_{t}|H|\psi_{t} \rangle$) after $\sigma_H^2$ reaches a relatively small value [$10^{-3}$ in the main text and $10^{-2}$ for Fig.~\ref{fig:DMRGS_SIMPS}(a)]. In contrast, the energy shift $\xi$ in SIMPS is fixed throughout the optimization. 
We emphasize that the update of the energy shift is crucial for extracting scars without \textit{a priori} knowledge of their precise locations in the energy spectrum:
In MBL systems, almost all the excited eigenstates satisfy the entanglement area law~\cite{bauer2013area,friesdorf2015many}. Due to the exponentially small energy spacings, in SIMPS the energy shift $\xi$ should be fixed in order to find \textit{all} the highly-excited eigenstates inside an energy window. 
In sharp contrast, in quantum many-body scarred systems, the energy spacings between adjacent scar tower states remain finite even in the thermodynamic limit [see Fig.~\ref{fig:Delta E_n}(a)]. Without a priori knowledge, the deviation between $\xi_0$ and the exact eigenenergies of scars can easily reach relatively large values.
Based on that, we let the energy shift $\xi$ slowly drift during the iterations, which enables our DMRG-S algorithm to converge to the desired scarred eigenenergies within large energy windows. As shown in Fig.~\ref{fig:DMRGS_SIMPS}(a), for the task of extracting the $E\approx -1.33$ scarred eigenstate in the PXP model, if we do not know its accurate eigenvalue in advance and choose the initial energy shift as $\xi_0=-1.6$, the SIMPS method with fixed $\xi$ (blue line) can not converge during the iterations (resulting in almost zero fidelity and wrong energy), whereas the DMRG-S method (red line) can successfully find the scarred eigenstate.

Second, for the shift-invert operator, in the DMRG-S algorithm we adopt $(H-\xi)^{-2}$ instead of $(H-\xi)^{-1}$. Note that different from the excited eigenstates of MBL systems, which satisfy the entanglement area law, most scar tower states near the middle of the spectrum possess logarithmic entanglement entropy. In the main text we need to take the maximum bond dimension as $\chi_{\rm max}=1200$ to compute the tower of scars in the PXP model of $L=80$. For such large bond dimensions, the efficiency of DMRG-S hinges on the positive semidefiniteness of the operator $(H-\xi)^2$, which enables us to adopt the conjugate gradient method to accelerate the solving of the linear equation with an iterative solver. We mention that even approximate solutions of the linear equation are sufficient for the convergence of the DMRG-S algorithm. 

Besides, through extensive numerical simulations, we find that compared with SIMPS, which adds one more $(H-\xi)$ operator on the right hand side of the equation and effectively uses $(H-\xi)^{-1}$ (see Fig. 1 of Ref.~\cite{yu2017finding}), the positive semidefinite $(H-\xi)^{-2}$ operator in DMRG-S significantly improves the robustness and efficiency of its convergence. The iterations become less frequently trapped in local minima of the optimization landscape. As demonstrated by the green line in Fig.~\ref{fig:DMRGS_SIMPS}(a) (to make a fair comparison we also update the energy shift $\xi$ in SIMPS), although the energy $E$ converges closely to the correct value, the fidelity exceeds $0.9$ in the first few iterations, then drops abruptly and finally gets trapped in a local minimum of $f \approx 0.7$.

In addition, we compare the performance of DMRG-S and SIMPS for the task of extracting scars with exact MPS representations. We take the exact  $\ket{\Gamma_{11}},\ket{\Gamma_{22}}$ scar states with $E=0$ in the PXP model~\cite{lin2019exact} as an example. 
For this task, we a priori know the eigenenergy of scars, and thus do not update the energy shift $\xi$ in DMRG-S. As shown in Fig.~\ref{fig:DMRGS_SIMPS}(b), we count the number of converged MPSs ($\sigma_H^2 <10^{-6}$) found by DMRG-S and SIMPS for different system sizes.
The DMRG-S algorithm turns out to be more efficient (see the inset) and have larger success probability in discovering the scars from the degenerate eigensubspace. 

We conclude that the designs in our DMRG-S algorithm make it well suitable and amenable for finding an isolated sub-volume-law entangled scar state in a sea of thermal eigenstates, without a prior knowledge of its precise eigenenergy. On the other hand, we should mention that the benefits brought by the update of energy shift and the usage of $(H-\xi)^{-2}$ are mainly specific to the task of extracting quantum many-body scars. 
For many-body localized systems, where the energy spacings between area-law entangled eigenstates are exponentially small, fixing the energy shift $\xi$ in SIMPS can ensure that the algorithm converges to an eigenstate in the vicinity of $\xi$ and finds all the excited eigenstates inside a small energy interval.

\ 
\ 

For the calculations of the scar-tower states in the PXP model, we need to make several adjustments upon the general DMRG-S algorithm, including the setting of special initial $\xi_0$ and the projection into the blockade constrained Hilbert space.
Note that we compute the tower of scars in the PXP model for the periodic boundary condition since their $\ket{Z_2}$ overlap and entanglement entropy constitute smoother functions of the energy, which are also easier to benchmark with ED for large system sizes by resolving the translation symmetry. During the iterations, we set the maximum bond dimension as $\chi_{\rm max}=1200$, while the open boundary cases roughly need $\sqrt{\chi_{\rm max}}$.

\textit{Setting of special initial $\xi_0$.--} The efficiency of the DMRG-S algorithm largely depends on the choice of the initial energy $\xi_0$ and initial state $|\psi_0 \rangle$. For the task of calculating the tower of scars in the PXP model, we set $\ket{\psi_0}=\ket{Z_2}$, which has predominant overlap with the $L+1$ scars within corresponding energy windows.
For the initialization of $\xi_0$, naively one can divide the whole possible energy interval into several windows and set $\xi_0$ as their medians (see the last part of this section). Another possible way is to use the energy obtained by the forward scattering approximation \cite{Turner2018weak}.
Here, according to our numerical experiences, we extrapolate the results in small system sizes to approximate the eigenenergy in larger systems:  In Fig. \ref{fig:Delta E_n}(b) we display the energy gap $\Delta E_n$ between adjacent scarred eigenstates as a function of $2n/L$ for $L=32,40,50,60,80$. We can observe that the curves for different system sizes overlap with each other, especially near the center of the spectrum. With the following ansatz function, we fit the curve of $L=32$, which can be obtained by exact diagonalization (ED), the forward scattering approximation or DMRG-S with random initial $\xi_0$. 
\begin{equation}
\Delta E_n=  \frac{b_0}{1+e^{b_1-b_2 (2n/L)}} + b_3.
\end{equation}
The fitting parameters are $(b_0,b_1,b_2,b_3) = (0.56,0.97, 5.19, 0.79)$. Now for large system sizes we can set $\xi_0$ as the approximate scarred eigenenergy
$E_n  = -\sum_{i = n}^{L/2-1}  \Delta E_i$ ($n=0,1,\cdots,L/2-1$, $E_{L/2}=0$, the other half could be obtained from the particle-hole symmetry $\{\prod_i Z_i, H_{PXP}\}=0$).
We stress that setting the special $\xi_0$ and the initial state as $\ket{\psi_0}=\ket{Z_2}$ are solely for this particular problem to get good convergence performance, and are not necessary for other general applications. In fact, in the last part of this section, we benchmark the robustness of DMRG-S by adopting random initial $\xi_0$ and random initial states.

\textit{Projection into the blockade constrained Hilbert space.--} Throughout the DMRG-S iterations, we expect that the optimized MPS always remain within the blockade constrained Hilbert space. Unfortunately it is not possible due to the approximate solving of the linear equation $\mathcal{A}_{t,\text{eff}}^{[i,i+1]} \psi_t^{[i,i+1]} = \tilde{\psi}_{t-1}^{[i,i+1]}$. 
Therefore, after we obtain the local tensor $\psi_t^{[i,i+1]}$, we additionally apply the two-qubit projector $1-n_i n_{i+1}$ [$n_i=(1+Z_i)/2$] on $\psi_t^{[i,i+1]}$  to project out the component $\ket{\uparrow\uparrow}_{i,i+1}$.  
\begin{figure}
\hspace*{-0.47\textwidth}
\includegraphics[width=0.47\linewidth]{ 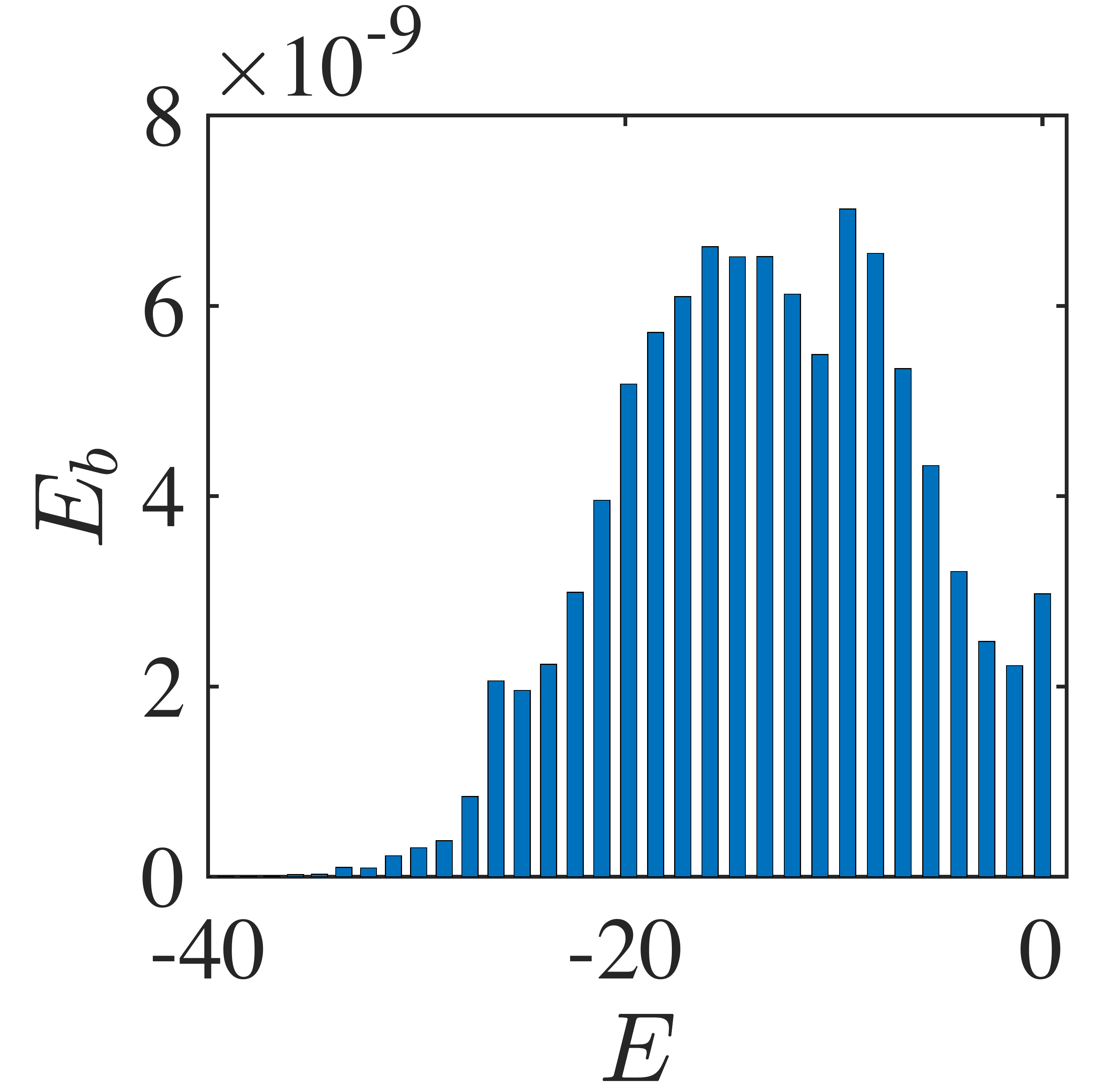} 
\caption{The Rydberg blockade energy $E_{b}= \bra{\Psi_n}(\sum_i n_i n_{i+1})\ket{\Psi_n}$ of the 41 scarred eigenstates with non-positive energy in the $L=80$ PXP model. 
}
\label{fig:E_b}
\end{figure}

Moreover, after finishing each iteration step $t$, we apply the projector $\prod_i (1- n_i n_{i+1})$ on the MPS $|\psi_t\rangle$ to further reduce the leakage out of the constrained Hilbert space. In Fig. \ref{fig:E_b}, we evaluate the Rydberg blockade energy $E_{b}= \bra{\Psi_n}(\sum_i n_i n_{i+1})\ket{\Psi_n}$ upon the tower of scars $\{\ket{\Psi_n}\}_{n=0}^{L}$ obtained by DMRG-S. We observe that the values for all the $E_b$ are of the order $10^{-8}$, confirming that these MPSs are indeed within the constrained Hilbert space.

\begin{figure}
\hspace*{-0.78\textwidth}
\includegraphics[width=0.78\linewidth]{ 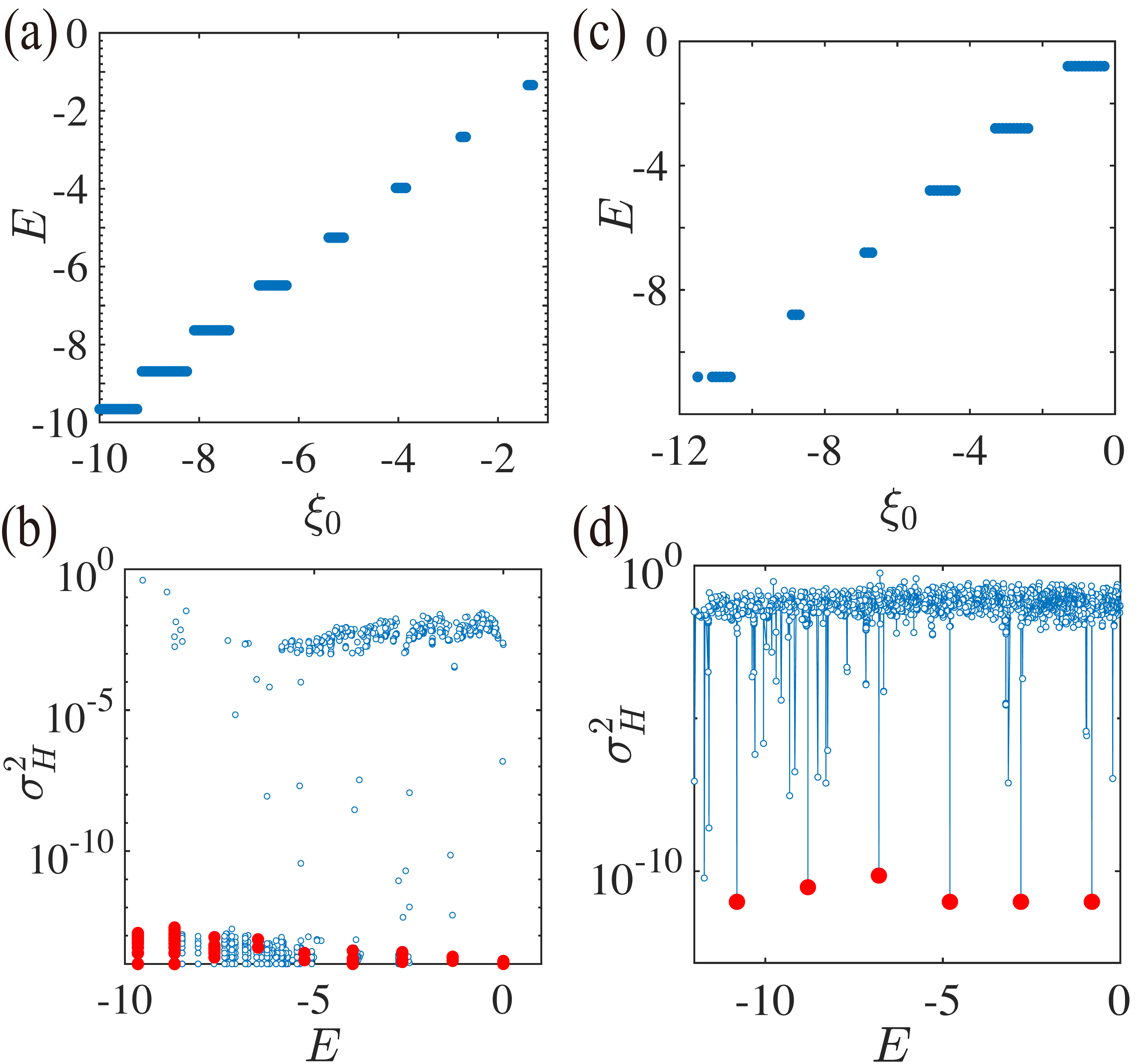} 
\caption{Testing the robustness of the DMRG-S algorithm for random initial $\xi_0$ and random initial states $|\psi_0\rangle$. 
(a) and (b) By using random $\xi_0$, the final converged energy $E$ as a function of $\xi_0$ for the $L=16$ PXP model and the $L=12$ spin-1 $XY$ model. The plateaus correspond to the eigenvalues of scarred eigenstates. $\xi_0\in[-10,0]$ with an interval $0.05$ and $\ket{\psi_0}=|Z_2\rangle$ for (a).
$\xi_0\in[-12,0]$ with an interval $0.1$ and $\ket{\psi_0}=|\phi_0\rangle$  for (b).
(c) and (d) By using random $\xi_0$ and random initial states $|\psi_0\rangle$, the energy variance $\sigma^2_H$ as a function of the converged energy $E$ for the $L=16$ PXP model and the $L=12$ spin-1 $XY$ model.   
For each $\xi_0$ [the same choices as (a) and (b)], we start the DMRG-S iterations from 10 random product states. The red dots mark the MPSs converged to the tower of scarred eigenstates.
}
\label{fig:robustness}
\end{figure}

\ 

Finally, we demonstrate the robustness of the DMRG-S algorithm by using random initial $\xi_0$ and random initial states $\ket{\psi_0}$. Besides the PXP model, we apply DMRG-S to extract the tower of scars in the 1D spin-1 $XY$ model \cite{Schecter2019Weak} 
\begin{equation}
H_\text{XY} = J \sum_{i} \left(S_{i}^{x} S_{i+1}^{x} + S_{i}^{y} S_{i+1}^{y}\right)
+ h \sum_{i} S_{i}^{z} + D \sum_{i}\left(S_{i}^{z}\right)^{2} + J_{3} \sum_{i}\left(S_{i}^{x} S_{i+3}^{x} + S_{i}^{y} S_{i+3}^{y}\right),
\end{equation} 
where $S_{i}^{\alpha}(\alpha=x, y, z)$ are spin-1 operators on the site $i$ and we take the parameters $(J,h,D,J_3) = (1 ,1 ,0.1,0.1)$. The spin-1 $XY$ model has been shown to be nonintegrable and host scarred eigenstates at $E_n = h(2n-L) + L D$ in OBC. Here the tower of scars support the perfect revival dynamics from the ground state $|\phi_0 \rangle$ of the staggered rhombic anisotropy Hamiltonian \cite{Schecter2019Weak}
\begin{equation}
H_{A}=\frac{1}{2} \sum_{i} (-1)^i \left[\left(S_{i}^{x}\right)^{2}-\left(S_{i}^{y}\right)^{2}\right].
\end{equation}

In Fig. \ref{fig:robustness}(a) and (b), we start the DMRG-S iterations by evenly picking up $\xi_0$ from the whole possible energy interval. The initial states remain as $\ket{\psi_0}=|Z_2\rangle$ for the $L=16$ PXP model (a) and $\ket{\psi_0}=|\phi_0\rangle$ for the $L=12$ spin-1 $XY$ model (b). The plateaus in $E$ correspond to the converged eigenvalues of scarred eigenstates, demonstrating that DMRG-S can successfully extract the particular scarred eigenstate within a finite range of $\xi_0$. The energy variance $\sigma_H^2$ for each converged MPS in (a) and (b) is less than $10^{-12}$.
Furthermore, in Fig. \ref{fig:robustness}(c) and (d), for each evenly picked $\xi_0$ we adopt 10 random product states as the initial state $\ket{\psi_0}$ and run the DMRG-S algorithm. We stress that under such conditions, we do not utilize any \textit{a priori} knowledge about the many-body systems. In (c) and (d), we can still find several converged MPSs with low energy variance marked by the red dots, whose energies exactly match up with the ED or analytical results.
Note that for the $L=16$ PXP model (c), since the system size is relatively small, there exist several converged MPSs with low energy variance not belonging to the tower of scars, which will disappear in larger system sizes. 
For all the computation in Fig. \ref{fig:robustness}, the maximum bond dimensions $\chi_{\rm max}$ are set to be $30\sim 60$ and the iteration steps are about $100$.
The above results once again demonstrate the power of our DMRG-S algorithm, which is confirmed to be a universal method applicable to generic many-body Hamiltonians.



\section{More numerical results}
In this section, we present more numerical results and analyses about the PXP model and the deformed PXP model, including the entanglement entropy of the scar-tower states, the eigenstate orders on the scar-tower states, the hybridizations between scarred and thermal eigenstates, and more observable dynamics within the scarred subspace.

\subsection{Entanglement Entropy and $|Z_2\rangle$ Overlap}

\begin{figure}
\hspace*{-0.75\textwidth}
\includegraphics[width=0.75\linewidth]{ 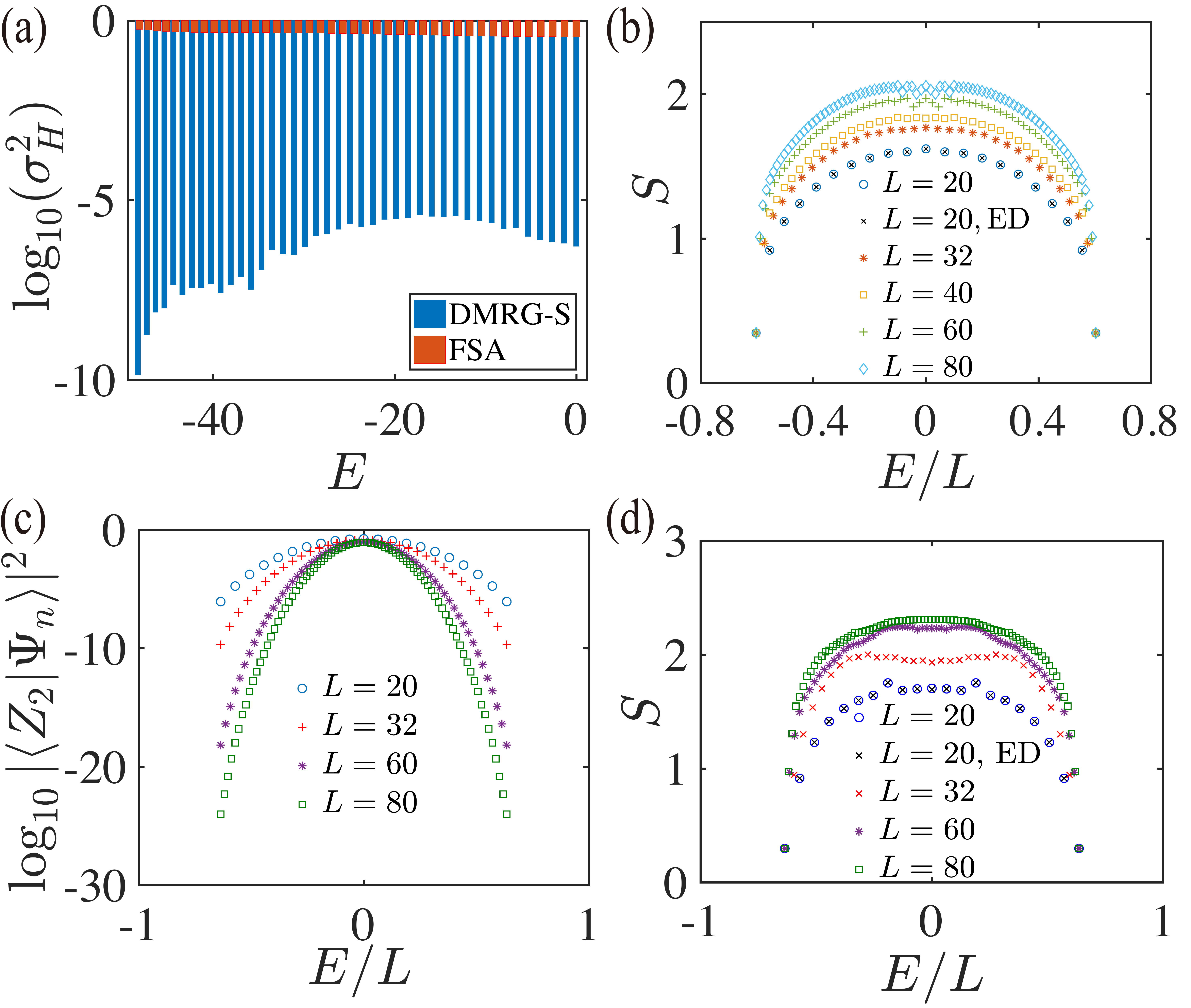} 
\caption{(a) Energy variance of the 41 scarred eigenstates with non-positive energy for the $L=80$ PXP model. The blue (red) bar results are obtained via DMRG-S (forward scattering approximation (FSA) \cite{Turner2018weak}). (b) Bipartite entanglement entropy $S$ of each scarred eigenstate in the PXP model for different system sizes. (c) Overlap between $|Z_2\rangle$ and each scarred eigenstate of the deformed PXP model for different system sizes. (d) Bipartite entanglement entropy of each scarred eigenstate in the deformed PXP model for different system sizes. For both models we use the periodic boundary condition.
}
\label{fig:entropy}
\end{figure}

In Fig. \ref{fig:entropy}(a), we compare the energy variance of the $41$ scarred eigenstates with non-positive energy for $L=80$ PXP model obtained via DMRG-S and the forward scattering approximation (FSA) \cite{Turner2018weak}. The average $\sigma_H^2$ obtained by DMRG-S is less than $10^{-6}$, much smaller than $10^{-1}$ by the FSA.
In Fig. 2(a) of the main text, we have used the $|Z_2\rangle$ overlap to validate the obtained MPSs for the PXP model. Here, in Fig. \ref{fig:entropy}(b) and (d) we further display their bipartite entanglement entropy $S$ for the PXP model and deformed PXP model. We can observe that: First, the energy and bipartite entanglement entropy of the ground state remain constant with increasing system sizes. The ground state is always gapped and satisfies the entanglement area law. 
Second, the entanglement entropy of other scar-tower states grows as the system size increases. And for a particular $L$, $S$ of the scarred eigenstates increases with the energy $E$ and reaches the maximum at $E=0$. 
Third, $S$ of each scar in the deformed PXP model is relatively larger than that of the corresponding scar in the PXP model.
Note that, as $L$ becomes larger, the curves of entanglement entropy collapse a little bit near $E=0$, which indicates that larger bond dimensions are needed to capture the entanglement property of scarred eigenstates near $E=0$.

We also display the $|Z_2\rangle$ overlap for the tower of scars in the deformed PXP model in Fig. \ref{fig:entropy}(c). Compared to Fig. 2(a) of the main text, we find that the $|Z_2\rangle$ overlap curves for both models tend to bend downward when the system size increases, whereas for the PXP model the curves shift downward more apparently. That mainly leads to the exponential decay of the total $|Z_2\rangle$ overlap illustrated in Fig. 2(b) of the main text.

For the calculations of the tower of scars in the deformed PXP model, we set the maximum bond dimension as $\chi_{\rm max}=400$, smaller that $\chi_{\rm max}=1200$ for the PXP model, mainly because the bond dimension of the MPO $(H-\xi_0)$ for the deformed PXP model is $14$, larger than $4$ for the PXP model.

\begin{figure}
\hspace*{-1.0\textwidth}
\includegraphics[width=1.0\linewidth]{ 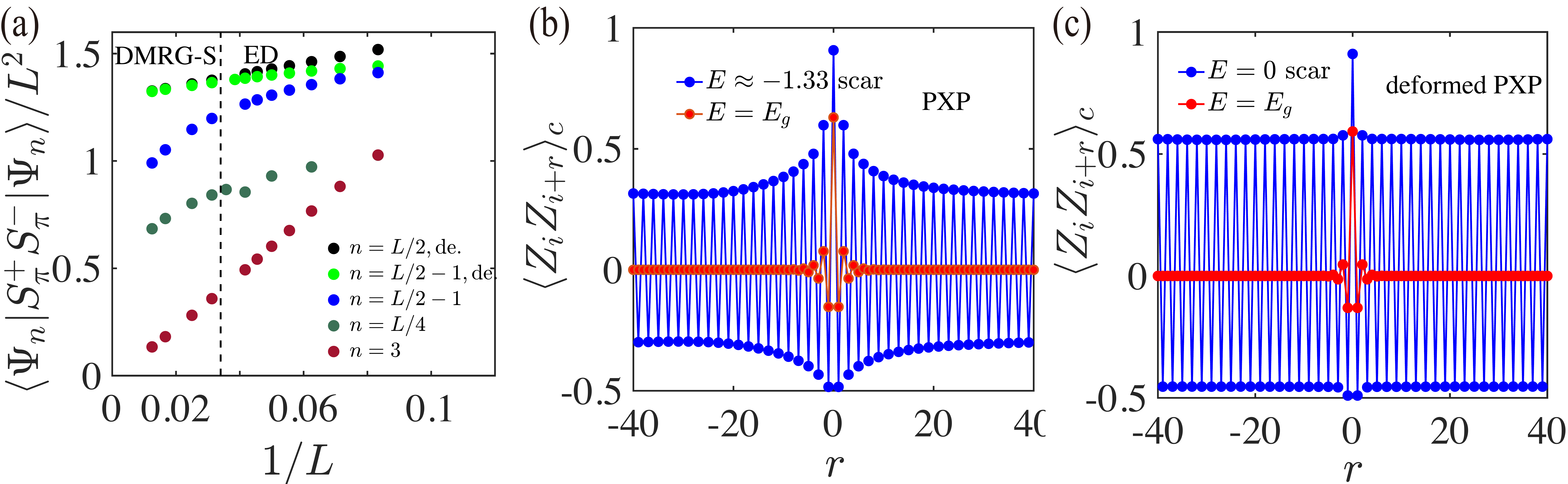} 
\caption{(a) Finite-size scaling for the $\pi$-magnon off-diagonal long-range order \cite{Iadecola2019Quantum} of scarred eigenstates in the (deformed) PXP model. The dashed vertical line separates data points obtained by DMRG-S or ED. (b) The connected correlations $\langle Z_i Z_{i+r} \rangle_{c}$ for the $E \approx -1.33$ scarred eigenstate and the ground state of the PXP model. $L=80$. (c) $\langle Z_i Z_{i+r} \rangle_{c}$ for the $E=0$ scar (at the middle of the spectrum) and the ground state of the deformed PXP model. $L=80$.}
\label{fig:zzcor}
\end{figure}

\subsection{Eigenstate Orders}
In Fig. \ref{fig:zzcor}(a), we demonstrate the off-diagonal long-range order $\langle S_{\pi}^{+} S_{\pi}^{-} \rangle/L^2$ upon the tower of scars $\{\ket{\Psi_n}\}_{n=0}^{L}$ to verify that the scarred eigenstates contain a finite density of $\pi$-magnons in the thermodynamic limit \cite{Iadecola2019Quantum}. The $\pi$-magnon creation and annihilation operators $S_\pi^{\pm}$ are defined as
\begin{equation}
   S_\pi^{\pm}=\frac{Z_\pi \mp 2 i Y_\pi}{2},
\end{equation}
where $Z_\pi=\sum_j (-1)^j Z_j$ and $Y_\pi=\sum_j (-1)^j P_{j-1} Y_j P_{j+1}$. By accessing system sizes beyond the ED regime , we confirm the existence of eigenstate orders on the scarred eigenstates, which are strictly forbidden by the eigenstate thermalization hypothesis at infinite temperature. Note that $\langle S_{\pi}^{+} S_{\pi}^{-} \rangle/L^2$ for the $n=L/2$ and $n=L/2-1$ scars of the deformed PXP model overlap with each other at large $L$, which is consistent with the $\pi$-magnon condensation interpretation since these two scars differ by only one $\pi$-magnon and have the same density in the thermodynamic limit.

Another natural consequence of the $\pi$-magnon condensation interpretation is that scarred eigenstates possess long-range connected correlations in space and time. In particular for the spatial correlation functions, $\lim_{r\to \infty} \langle Z_i Z_{i+r} \rangle_\text{c,scar} \sim (-1)^r \times \text{const}$, where $\langle Z_i Z_{i+r} \rangle_c = \langle Z_i Z_{i+r} \rangle - \langle Z_i \rangle \langle Z_{i+r} \rangle$ \cite{Iadecola2019Quantum}. 
In Fig. \ref{fig:zzcor}(b) and (c) we demonstrate the spatial eigenstate orders on the scarred eigenstates of the PXP and deformed PXP model, with the system size $L=80$ beyond the ED regime. The spatial correlation functions indeed exhibit long-range orders for the highly excited scarred eigenstates, whereas vanish on the ground state due to its zero $\pi$-magnon density.

\subsection{Hybridizations between Scarred and Thermal Eigenstates}
\begin{figure}
\hspace*{-1.0\textwidth}
\includegraphics[width=1.0\linewidth]{ 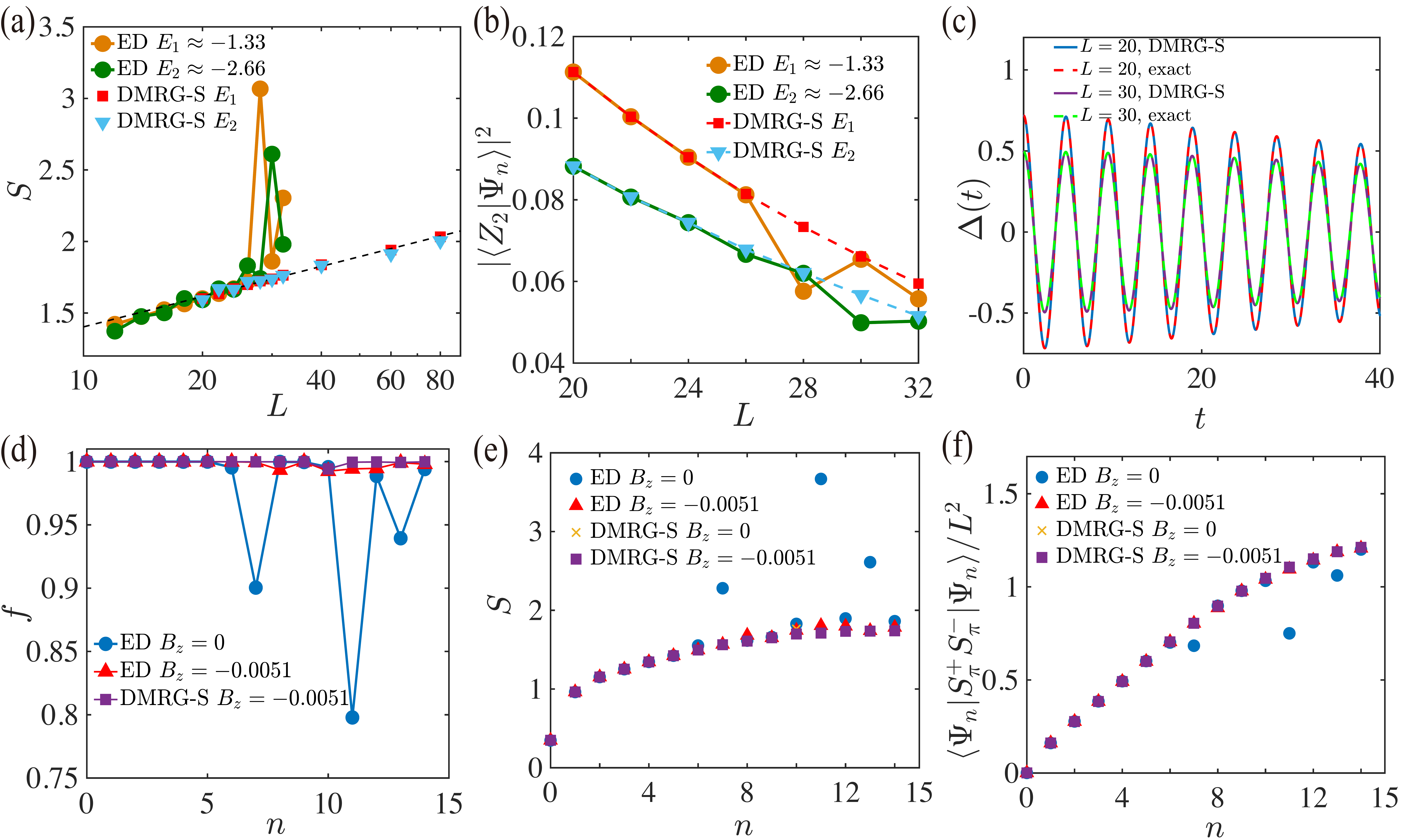} 
\caption{Hybridizations between scarred and thermal eigenstates in the PXP model. (a) The bipartite entanglement entropy $S$ of the $E_1\approx -1.33$ and $E_2 \approx -2.66$ scars obtained by ED and DMRG-S for different system sizes. The dashed line is the logarithmic fitting with $S=0.70\log_{10}(L) + 0.70$. (b) The overlap between $|Z_2\rangle$ and the $E_1$, $E_2$ scars obtained by ED and DMRG-S for different system sizes. (c) The $\Delta(t)$ dynamics computed by using the DMRG-S eigenenergies $\{E_n\}_{n=0}^L$ and by exact Hamiltonian evolution. (d) The fidelity $f$ between the scars for the $B^z=0$ case obtained by DMRG-S and the scars for the other three cases ($B^z=0$ obtained by ED, $B^z=-5.1\times 10^{-3}$ obtained by ED and $B^z=-5.1\times 10^{-3}$ obtained by DMRG-S). Here $L=30$ and we plot the results for scars with negative energy ($n=0,1,\cdots,14$). (e) The bipartite entanglement entropy of the scar-tower states for the above four cases. (f) The $\pi$-magnon off-diagonal long-range order of scars for the above four cases.
}
\label{fig:hybridization}
\end{figure}

For the PXP model, previous diagonalization calculations of the entanglement entropy and the $\pi$-magnon off-diagonal long-range order for the scar-tower states sometimes show  nonmonotonic scaling behaviors with the system size (see the outliers in Fig. 8 of Ref.~\cite{Turner2018quantum} and Fig. 7 of Ref.~\cite{Iadecola2019Quantum}). These outliers can be attributed to accidental hybridizations, or eigenstate resonances, between the scarred eigenstates and thermal eigenstates nearby in energy. In Fig.~\ref{fig:hybridization} we carefully address that these hybridized scarred eigenstates will not affect the physical conclusions drawn in the main text.

First, similar to Fig.~8 of Ref.~\cite{Turner2018quantum}, in Fig.~\ref{fig:hybridization}(a) we compare the bipartite entanglement entropy of the $E_1\approx -1.33$ ($n=L/2-1$) and $E_2 \approx -2.66$ ($n=L/2-2$) scars obtained by ED and DMRG-S for different system sizes. When the hybridizations happen, the DMRG-S method captures the low-entanglement part of hybridized scarred eigenstates. This is intrinsically restricted by the filter property of the maximum bond dimension. 

Nevertheless, the above low-entanglement bias does not affect our conclusions about the N\'eel state revivals in the PXP model: For any accessible system sizes by ED ($L\leq 32$), we observe that the $\ket{Z_2}$ overlaps of hybridized scarred eigenstates are all smaller than the smooth values estimated by DMRG-S [Fig.~\ref{fig:hybridization}(b)]. So the total $\ket{Z_2}$ overlap of all the scar-tower states in the PXP model still decays exponentially with the system size even considering the hybridizations.

As for the observable dynamics computed in the main text, recall that we evaluated the dynamics of the staggered magnetization density within the scarred subspace constructed by DMRG-S: $\Delta (t) = \langle Z_2 | \mathbb{P}  e^{i H t}  \Delta e^{-i H t} \mathbb{P} | Z_2 \rangle \approx \sum_{n,m=0}^L e^{i(E_n - E_m)t}$ $ \langle Z_2 | \Psi_n \rangle \langle \Psi_n | \Delta | \Psi_m \rangle \langle \Psi_m | Z_2 \rangle$, where $\mathbb P = \sum^L_{n=0}\ket{\Psi_n}\bra{\Psi_n}$, $\{E_n\}_{n=0}^L$ and $\{\ket{\Psi_n}\}_{n=0}^L$ are scarred eigenenergies and eigenstates obtained via DMRG-S. In Fig. 2(e) of the main text and Fig.~\ref{fig:hybridization}(c), we explicitly benchmark the $\Delta(t)$ dynamics calculated by using DMRG-S eigenenergies and by exact Hamiltonian evolution, for the $L=20$ (without hybridized scars) and $L=28, 30$ (with hybridized scars) PXP model, all of which show good agreement. Since by its definition the hybridization can not alter the eigenenergies of scars drastically, as long as the eigenenergy variances are small enough for the DMRG-S states ($ e^{-i H t} \ket{\Psi_n} \approx e^{-i E_n t} \ket{\Psi_n} $), $\Delta(t)$ computed by using the DMRG-S eigenenergies should agree with the exact results.

Besides, we demonstrate that the hybridizations between scarred and thermal eigenstates are just some accidental effects. The hybridizations are not robust and can be removed by some tiny perturbations. 
As shown in Fig.~\ref{fig:hybridization}(d), we compute the state fidelity between the scars obtained by ED and DMRG-S for $L=30$, where there are three hybridized outliers (blue dots). The fidelities of normal scars are almost one (with typical infidelities about $10^{-4}$), while the fidelities for the three hybridized outliers are around $0.8$ to $0.95$. After we add a tiny $z$-direction magnetic field term $B^z \sum_{i=1}^L Z_i$ to the PXP Hamiltonian, the three outliers disappear (red dots), whereas the scars obtained by DMRG-S remain stable (purple dots). 

We also benchmark the entanglement entropy [Fig.~\ref{fig:hybridization}(e)] and the $\pi$-magnon off-diagonal long-range order [Fig.~\ref{fig:hybridization}(f)] of scars for the unperturbed and slightly perturbed cases by ED or DMRG-S, which display consistent results with Fig.~\ref{fig:hybridization}(d). The states extracted by DMRG-S are more physical and more consistent with the defining features of many-body scars (robust to perturbations, smaller entanglement entropy, larger $\ket{Z_2}$ overlap, more like a $\pi$-magnon condensate). Note that we choose the value $B^z=-5.1\times 10^{-3}$ because it shows good performance to remove hybridizations for all the three outliers of $L=30$. Other $B^z$ values and other types of perturbations could also mitigate the hybridization effect to some extent.

\begin{figure}
\hspace*{-0.8\textwidth}
\includegraphics[width=0.8\linewidth]{ 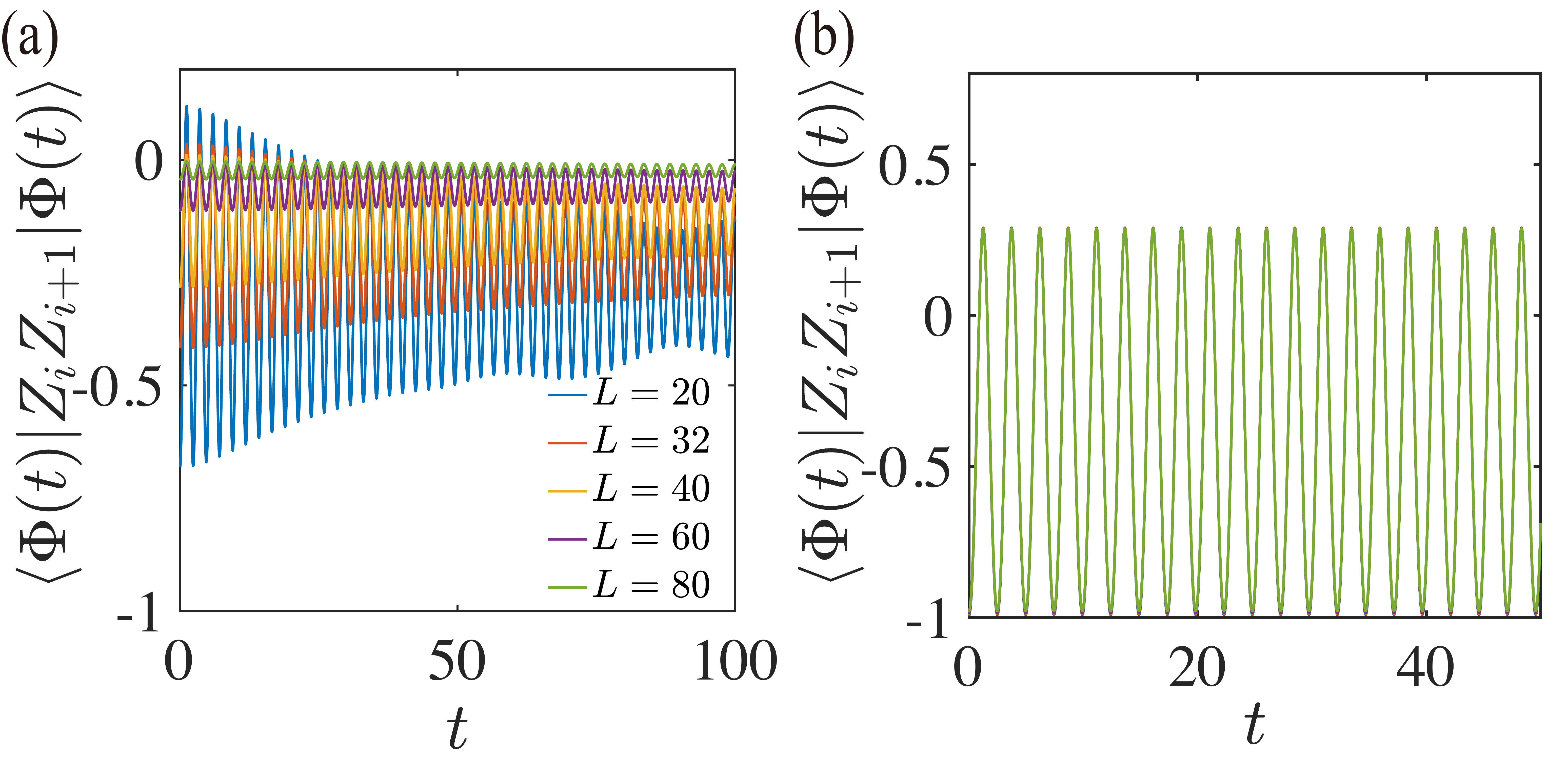} 
\caption{Dynamics of the correlation function $\langle Z_i Z_{i+1} \rangle$ evaluated upon $|Z_2 \rangle$ projected into the scarred subspace constructed by DMRG-S ($\mathbb{P} = \sum_n |\Psi_n\rangle \langle \Psi_n|$), for the PXP model (a) and the deformed PXP model (b). $|\Phi(t)\rangle = e^{-iHt} \mathbb{P} |Z_2\rangle$, $i=1$.
}
\label{fig:obs_dyn}
\end{figure}

\subsection{Observable Dynamics within the Scarred Subspace}
In addition to the $\Delta(t)$ dynamics shown in Fig. 2(c) and (d) of the main text, in Fig. \ref{fig:obs_dyn} we calculate observable dynamics of the $Z_i Z_{i+1}$ correlation evaluated upon $|Z_2 \rangle$ projected into the scarred subspace constructed by DMRG-S ($\mathbb{P} = \sum_n |\Psi_n\rangle \langle \Psi_n|$). $|\Phi(t)\rangle = e^{-i H t} \mathbb{P} |Z_2\rangle$ and we apply the DMRG-S eigenenergies to compute its dynamics. Similar to $\Delta(t)$, we can observe an obvious envelope decay of the oscillations for the PXP model as the system size increases, while the oscillations for the deformed PXP model remain nearly perfect when $L$ increases.

\begin{figure}
\hspace*{-1.0\textwidth}
\includegraphics[width=1.0\linewidth]{ 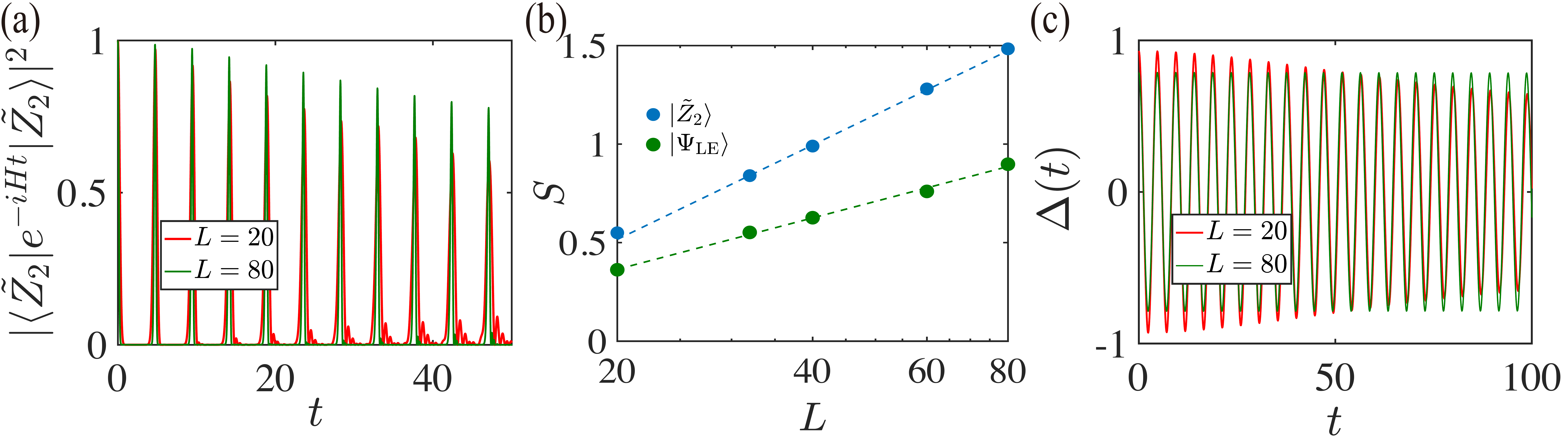} 
\caption{(a) Loschmidt echo of the deformed $Z_2$ state $\ket{\tilde{Z_2}}$ for the PXP model. (b) The bipartite entanglement entropy of $\ket{\tilde{Z_2}}$ and $|\Psi_{\rm LE}\rangle$ as a function of the system size. The dashed lines are logarithmic fittings with $S =1.66\log_{10}(L) - 1.7$ for $\ket{\tilde{Z_2}}$ and $S = 0.87\log_{10}(L) - 0.76$ for $|\Psi_{\rm LE}\rangle$. (c) Observable dynamics $\Delta(t)$ of the ``least entangled" state $|\Psi_{\rm LE}\rangle$ for the PXP model.
}
\label{fig:tilde_Z2}
\end{figure}

One interesting observation from Fig. 2(c) of the main text and Fig.~\ref{fig:obs_dyn}(a) is that for the PXP model when the system size increases, 
the envelope of oscillations, despite becoming smaller overall, \textit{decays slower} with the evolution time. This phenomenon indicates that the stability of oscillations actually increases with the system size. As mentioned in the main text, we define the deformed $Z_2$ state as $\ket{\tilde{Z_2}} = \mathbb{P} \ket{Z_2} / \sqrt{\langle Z_2 | \mathbb{P} | Z_2 \rangle}$. In addition to the observable dynamics $\Delta(t)$ shown in Fig. 2(e) of the main text, we calculate the Loschmidt echo $|\bra{\tilde{Z_2}}  e^{-i H t} \ket{\tilde{Z_2}}|^2$ by using the DMRG-S eigenenergies. As shown in Fig. \ref{fig:tilde_Z2}(a), recurrence peaks of the Loschmidt echo are even higher for larger system sizes, implying that periodic revivals of $\ket{\tilde{Z_2}}$ are stable in the thermodynamic limit. Note that compared to the product state $\ket{Z_2}$, $\ket{\tilde{Z_2}}$ still possesses modest entanglement, which scales logarithmically with the system size [Fig.~\ref{fig:tilde_Z2}(b)]. 

Moreover, starting from $\ket{\tilde{Z_2}}$, we utilize the algorithm in Ref.~\cite{karle2021area,Reuvers2018Algorithm} (exponentiating the singular values and projecting back to the scarred subspace) to reduce its bipartite entanglement entropy and find the ``least entangled" state $|\Psi_{\rm LE}\rangle$. During the optimization we require $|\Psi_{\rm LE}\rangle$ to have symmetric $\ket{Z_2}$ overlap $|\langle Z_2 | \Psi_n \rangle|^2$ with respect to $E=0$. As shown in Fig.~\ref{fig:tilde_Z2}(b), the entanglement entropy of the optimized $|\Psi_{\rm LE}\rangle$ also scales logarithmically with the system size, yet with a smaller slope than that of $\ket{\tilde{Z_2}}$. Oscillations in the observable dynamics $\langle \Psi_{\rm LE} | e^{i H t} \Delta e^{-i H t} | \Psi_{\rm LE} \rangle$ also become more stable for larger system sizes [Fig.~\ref{fig:tilde_Z2}(c)].

These results imply that the stable periodic revival dynamics \textit{inhere within} the scarred subspace of the PXP Hamiltonian, while the exponential decay of $\langle Z_2 | \mathbb{P} | Z_2 \rangle$ leads to the envelope decay observed previously. One needs to take other modestly entangled initial states like $\ket{\tilde{Z_2}}$ or $|\Psi_{\rm LE}\rangle$ to exhibit that.
The above analyses could provide another perspective to explain why quasilocal deformations of the original PXP Hamiltonian are able to make the revivals virtually perfect \cite{Choi2019emergent}.

\end{document}